\documentclass[aps,pre,groupedaddress,superscriptaddress,showkeys,showpacs,amsmath]{revtex4-1}
\usepackage{amssymb}
\usepackage{amsmath}
\usepackage{graphicx}
\usepackage{amsbsy}
\usepackage{bm}
\usepackage{color}
\usepackage{gensymb}
\usepackage{caption}
\usepackage{subcaption}
\usepackage[space]{grffile}
\definecolor{cream}{RGB}{222,217,201}
\usepackage{siunitx}
\DeclareSIUnit[number-unit-product = {\,}]
\cal{cal}
\DeclareSIUnit\kcal{\kilo\cal}
\DeclareSIUnit\kcal{\kilo\joule\per\mole}
\DeclareSIUnit\molar{\mole\per\cubic\deci\metre}
\DeclareSIUnit\Molar{\textsc{m}}
\usepackage{multirow}

\begin{document}

\title{ Local symmetry determines the phases of linear chains: a simple model for the self-assembly of peptides }

\author{Tatjana \v{S}krbi\'{c}} 
\email{tatjana.skrbic@unive.it}
\affiliation{Dipartimento di Scienze Molecolari e Nanosistemi, Universit\`{a} Ca' Foscari di Venezia
Campus Scientifico, Edificio Alfa, via Torino 155,30170 Venezia Mestre, Italy}
\affiliation{Department of Physics and Institute for Theoretical Science, 1274 University of Oregon, 
  Eugene, OR 97403-1274, USA}

\author{Trinh X. Hoang}
\email{hoang@iop.vast.ac.vn}
\affiliation{Center for Computational Physics, Institute of Physics, Vietnam Academy of Science and Technology,
  10 Dao Tan, Ba Dinh, Hanoi, Vietnam}

\author{Amos Maritan}
\email{amos.maritan@pd.infn.it}
\affiliation{Dipartimento di Fisica e Astronomia, Universit\`a di Padova, and INFN, via Marzolo 8,I-35131 Padova, Italy}

\author{Jayanth R. Banavar}
\email{banavar@uoregon.edu}
\affiliation{Department of Physics and Institute for Theoretical Science, 1274 University of Oregon, 
Eugene, OR 97403-1274, USA}

\author{Achille Giacometti} 
\email{achille.giacometti@unive.it}
\affiliation{Dipartimento di Scienze Molecolari e Nanosistemi, Universit\`{a} Ca' Foscari di Venezia
Campus Scientifico, Edificio Alfa, via Torino 155,30170 Venezia Mestre, Italy}

\date{\today}

\begin{abstract}
  We discuss the relation between the emergence of new phases with broken symmetry within the framework of simple models of biopolymers. We start with a classic model for a chain molecule of spherical beads tethered together, with the steric constraint that non-consecutive beads cannot overlap, and with a pairwise attractive square well potential accounting for the hydrophobic effect and promoting compaction. We then discuss the consequences of the successive breaking of spurious symmetries. First, we allow the partial interpenetration of consecutive beads. In addition to the standard high temperature coil phase and the low temperature collapsed phase, this results in a new class of marginally compact ground states comprising conformations reminiscent of $\alpha$-helices and $\beta$-sheets, the building blocks of the native states of globular proteins. We then discuss the effect of a further symmetry breaking of the cylindrical symmetry on attaching a side-sphere to the backbone beads along the negative normal of the  chain, to mimic the presence of side chains in real proteins. This leads to the emergence of a novel phase within the previously obtained marginally compact phase, with the appearance of more complex secondary structure assemblies.  The potential importance of this new phase in the \textit{de novo} design of self-assembled peptides is highlighted.
\end{abstract}
\maketitle

\section{Introduction}
\label{sec:introduction}
We unify two themes in this paper, one from statistical physics and the other in the life sciences. The notion of phases of matter plays a major role in statistical physics. Even the very simplest many body system of hard spheres exhibits two phases on varying the packing fraction. One obtains a fluid phase at low packing fraction and a crystalline phase at sufficiently high packing fraction. This is a purely entropic effect, which takes into account the number of ways a system of hard spheres can be arranged while ensuring that the hard spheres do not overlap. On adding a short range attraction between the spheres, the fluid phase splits into two phases -- the liquid phase and the vapor phase \cite{Chaikin00,Hansen06}. 

Symmetry plays a key role in determining the nature of the ordered phase. Consider replacing a packing of spheres with a packing of anisotropic objects such as pencils or banana shaped objects. The spherical symmetry of the constituent objects has been broken by hand and one can contemplate a situation in which the translational ordering need not occur in all three directions simultaneously. Furthermore, one may have the possibility of orientational ordering without any accompanying translational order \cite{Chaikin00}. For example, the center of masses of the pencils can be disordered, yet the pencils may all roughly be oriented along the same axis. 

The notion of phases and singularities associated with phase transitions are all well-defined only in the thermodynamic limit for an infinite sized system. For a finite sized system, the behavior typically mirrors that of an infinite system but with the singularities being rounded out. 

The second theme that our work touches upon is the behavior of relatively short chain molecules of amino acids (there are twenty types of naturally occurring amino acids), proteins \cite{Cantor80,Finkelstein16}. Proteins are amazing molecular machines that do the work in a living cell. The behavior of proteins is largely governed by their native state structure or loosely speaking their ground state geometry. Protein native states are made up of emergent building blocks of tightly wound helices and almost planar zig-zagging strands forming $\beta$-sheets. Both helices and sheets are stabilized by hydrogen bonds as first shown by Pauling and co-workers many decades ago \cite{Pauling51,Pauling51b}. Furthermore, proteins are able to change their geometry due to external influences such as binding to ligands or other proteins or signalling molecules. Similar proteins are present in all living cells and thus one might wonder whether their very special attributes arise from their native state structures lying in a novel phase of matter that exists for relatively short chain molecules and which comprise the common geometrical attributes of all proteins. Our goal is to identify such a phase, which, we will show, exists independent of the details of quantum chemistry and amino acid specificity. 

We begin with a simple classical model of a short chain molecule and, guided by symmetry considerations, study the nature of the ground states using extensive computer simulations. Guided by our experience in statistical physics, we monitor the nature of the ground states on successively breaking spurious symmetries. Without any additional input from quantum chemistry or amino acid heterogeneity, we identify a novel phase of matter, which meets the requirements we seek. We then study the behavior of the phase diagram as a function of temperature to ensure that the novel phase is still viable at non-zero temperatures. We then conclude with a careful comparison of the structures in the novel phase and assembled protein native state structures. Note that the notion of a phase has to be treated with appropriate caution for a finite sized system as alluded to above.

Our work has several consequences. First, it provides some new insights into the behavior of simple models of self-avoidance of chain molecules subject to an attractive self-interaction. Second, it highlights the all-important role of symmetry in determining the nature of the ground states. Finally, it has potential applications in the design and control of the self-assembly process of peptides yielding nanostructures with prescribed properties \cite{Huang16,Ljubetic17,Shen18,Li19,Bera19}. 

A standard model of a homopolymer chain comprises $N$ spherical beads of diameter $\sigma$ tethered into a chain of total length $L$. Consecutive beads are kept at a fixed distance $b=\sigma$ to represent covalent bonds ($L = (N-1) b$), and non-consecutive beads are not allowed to overlap and account for excluded volume effects \cite{deGennes79,Khokhlov02,Rubinstein03}.  At this level of description, solvent effects are incorporated by including a short range square-well attraction between non-consecutive beads, so that these beads prefer to stay close to one another, as for a chain in a bad solvent. The phase diagram is well known and has been studied by many different groups using various methods \cite{deGennes79,Taylor09b,Skrbic16a}. It has a high temperature swollen phase, where the chain is in a relatively open stretched conformation dominated by entropy. At low temperatures, the energy dominates resulting in a compact phase, where the number of attractive contacts is high. For sufficiently long chains, the ground state has a crystalline structure, usually with FCC or HCP local symmetry \cite{Taylor09b}. 

There is an intrinsic problem with the symmetry of the model. Untethered spheres look the same from any direction -- they are isotropic. The act of being tethered together in a chain causes the spherical symmetry of monomers to conflict with the fact that there is a tangent direction at each sphere location, which describes the direction of the chain. Thus a spherical geometry of the tethered objects is not compatible with the symmetry associated with a chain.

Proteins are inherently different from synthetic polymers. A polypeptide chain is formed by a sequence of amino acids characteristic of each protein, with repeated units formed by several atoms rather than a single monomer entity \cite{Cantor80,Finkelstein16}. While, at the simplest level, this could be accounted for within a single bead representation, some known structural aspects do not fit the square-well bead model described above.  Consider the simplest example of a poly(GLY) chain, where all amino acids are glycines (GLY) having just a single hydrogen atom as a side chain. These beads would correspond to a van der Waals sphere associated with GLY and having a diameter of the order of $5$ {\AA}, whereas the distance between the centers of consecutive beads, the C$_{\alpha}$- C$_{{\alpha}^{\prime}}$ distance in protein language, is known from crystallographic measurements to be approximately $3.81 ${\AA}. This alone already suggests that $b<\sigma$, unlike the previous assumption of $b=\sigma$. Real proteins are however not poly(GLY) and the chain includes amino acids of 20 different types, differing from each other because of distinctive side chains. Naturally occurring side chains have, in general, different sizes (GLY is very small, Tryptophan TRP is, on the contrary, bulky), as well as different chemical properties. The side chains are typically oriented in a direction perpendicular to the backbone chain and provide additional steric constraints and chemical attributes.

Motivated by these features, a model called a ``Thick Chain'' (TC) or `` Tube'' model was proposed some years ago \cite{Maritan00}. In this model, the chain of spherical beads was replaced by a flexible (able to be bent locally below a certain threshold with no energy cost), continuum (with no discrete granularity) tube, with a diameter given by $2\Delta$. Here, $\Delta$ is defined to be the ``thickness'' of the tube \cite{Gonzalez99} and encapsulates the ability to house side chains within it, with a larger $\Delta$ allowing for bigger side chains. Of course, a homopolymer would be represented by a tube of uniform thickness $\Delta$. The transition from a chain of spheres to a continuum tubular object has two important consequences for the conformational statistics of the chain. As in the case of the discrete chain, different parts of the tube cannot overlap. In addition, the tube cannot be bent too severely locally with the constraint that the local radius of curvature is no smaller than the thickness $\Delta$. 

An important aspect of the TC model is related to its symmetry. In a spherical-bead model, any given monomer is spherically symmetric. In the TC model, the tube axis provides a preferred directionality thus breaking the original local spherical symmetry in favor of a cylindrical symmetry. As in liquid crystals \cite{Chaikin00}, this symmetry breaking can result in new phases, in addition to the conventional coil (swollen) and the globular (or crystalline) phases. 

While the picture of a continuum tube is very handy from the conceptual point of view, it cannot be used in practical terms because a discretization is always necessary. It turns out that the continuum tube can be recast in terms of a discrete chain with a suitable three-body potential \cite{Gonzalez99,Stasiak00,Maritan00}. The three body potential is however very costly from the computational view point, and several studies \cite{Clementi98,Magee07,Banavar09,Coluzza11,Skrbic16a,Skrbic16b,Werlich17} have suggested the alternative route of allowing partial interpenetration between consecutive monomer beads, prompted by the structural motivations alluded to earlier. The $b<\sigma$ condition provides an entropic constraint rather similar to that in the TC model. This solution, however, does not give one the possibility of tuning the thickness, as in the thick chain model. One possible way of approximately accomplishing this is to add a necklace of additional spherical beads (side chain beads) surrounding the main chain bead in a plane perpendicular to the chain axis and tangent to each backbone sphere, akin to fixed satellites. Upon varying the number $N_{sc}$ and the diameter $\sigma_{sc}$ of the side spheres, an effect similar to that of the thick chain model can be achieved.  

Here, we build upon these ideas by studying the effect of adding a single side chain located at a specific position in the ring, thus further reducing the uniaxial (cylindrical) symmetry to biaxial. We will show that this leads to an additional sub-phase with highly unconventional properties that will be the focus of the present study.  A preliminary report of this novel phase has been presented before \cite{Skrbic19,Rose19a} -- the present study provides a complete analysis including studies of the behaviour at non-zero temperatures.

The rest of the paper is organized as follows. Section \ref{sec:math} will recall the mathematical formalism necessary for a proper descriptions of the models. Section \ref{sec:results} include detailed results and discussions. Finally, Section \ref{sec:conclusions} will summarize the results and discuss some future perspectives. 

\section{Model and Methods}
\label{sec:math}
\subsection{The model}
\label{subsec:model}
\begin{figure}[htpb]
\centering
 \captionsetup{justification=raggedright,width=\linewidth}
  \begin{subfigure}{6cm}
     \includegraphics[width=.8\linewidth]{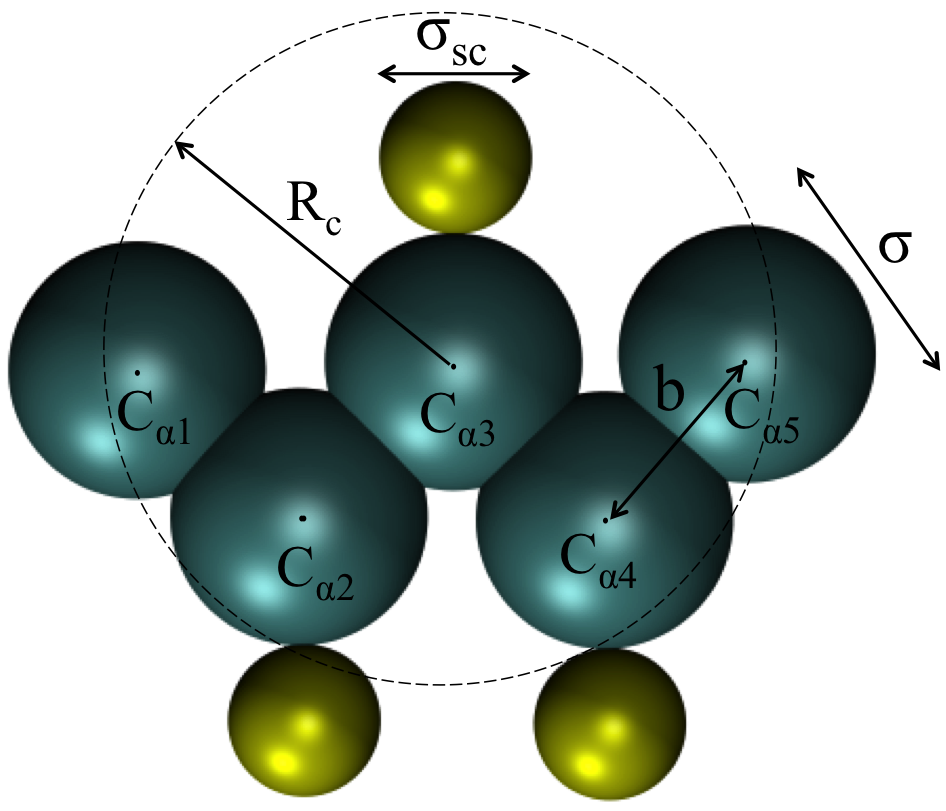}
    \caption{}\label{fig:fig1a}
  \end{subfigure}
  \begin{subfigure}{6cm}
     \includegraphics[width=.8\linewidth]{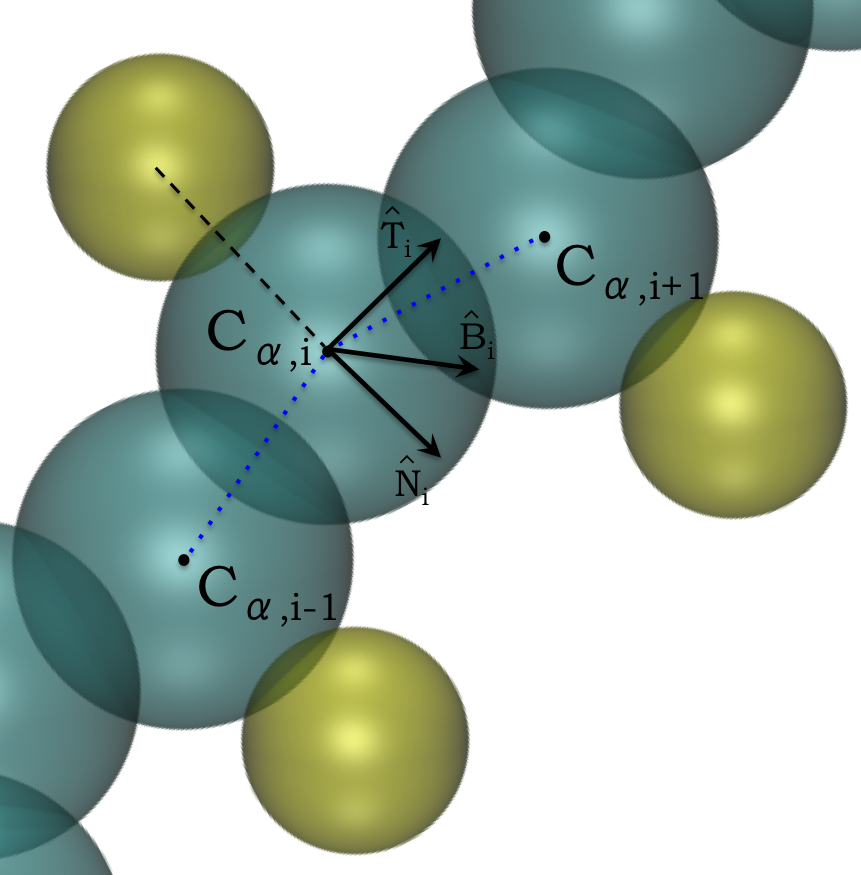}
    \caption{}\label{fig:fig1b}
  \end{subfigure}
  \caption{(a) Chain model. Each main chain sphere (cyan) has diameter $\sigma$. The side sphere (yellow) is in the negative normal direction and has a diameter $\sigma_{sc}$. The distance between successive main chain spheres is $b\le \sigma$ -- consecutive spheres can, in general, partially overlap. Non-consecutive main chain spheres experience a short range attractive constant potential if their separation is within the range of the attraction $R_c$. (b) Side sphere positions in Frenet frame. 
  \label{fig:fig1}}
\end{figure}

Our model, inspired by past studies \cite{Magee07,Banavar09,Skrbic16a,Skrbic16b} is displayed in Figure \ref{fig:fig1}. It consists of a chain of $N$ identical tethered spherical  beads of diameter $\sigma$, each representing the backbone of an amino acid centered at each C$_{\alpha}$, and having a nearest neighbour (along the chain) distance equal to $b \le \sigma$. Figure \ref{fig:fig1a} shows the case $N=5$. To each of the $N-2$ internal beads, a second bead of
diameter $\sigma_{sc}$ is attached tangent to the backbone bead and located along the negative normal $\widehat{\mathbf{N}}$ direction of the Frenet frame $\{\widehat{\mathbf{T}},\widehat{\mathbf{N}},\widehat{\mathbf{B}}\}$ \cite{Coxeter69} (see Figure \ref{fig:fig1b}). In addition to excluded volume involving all beads (backbone-backbone, backbone-side chain, and side chain-side chain), a constant short range attraction of strength $\epsilon$ and range $R_c$ is imposed between the main chain beads as depicted in Figure \ref{fig:fig1a}. Overall then, our model is characterized by three parameters: the inter-bead distance $b/\sigma$, the size of the side chain $\sigma_{sc}/\sigma$, and the range of attraction $R_c/\sigma$. Realistic values for these three parameters in proteins are $b=3.81$ {\AA}, $\sigma_{sc}=2.5$ {\AA}, and $R_c=6$ {\AA}; a realistic value for the diameter of the backbone bead is $\sigma=5$ {\AA} corresponding to the diameter of the van der Waals sphere associated with Glycine (GLY).  

\begin{figure}[htpb]
  \centering
  \captionsetup{justification=raggedright,width=\linewidth}
  \includegraphics[width=0.7\linewidth]{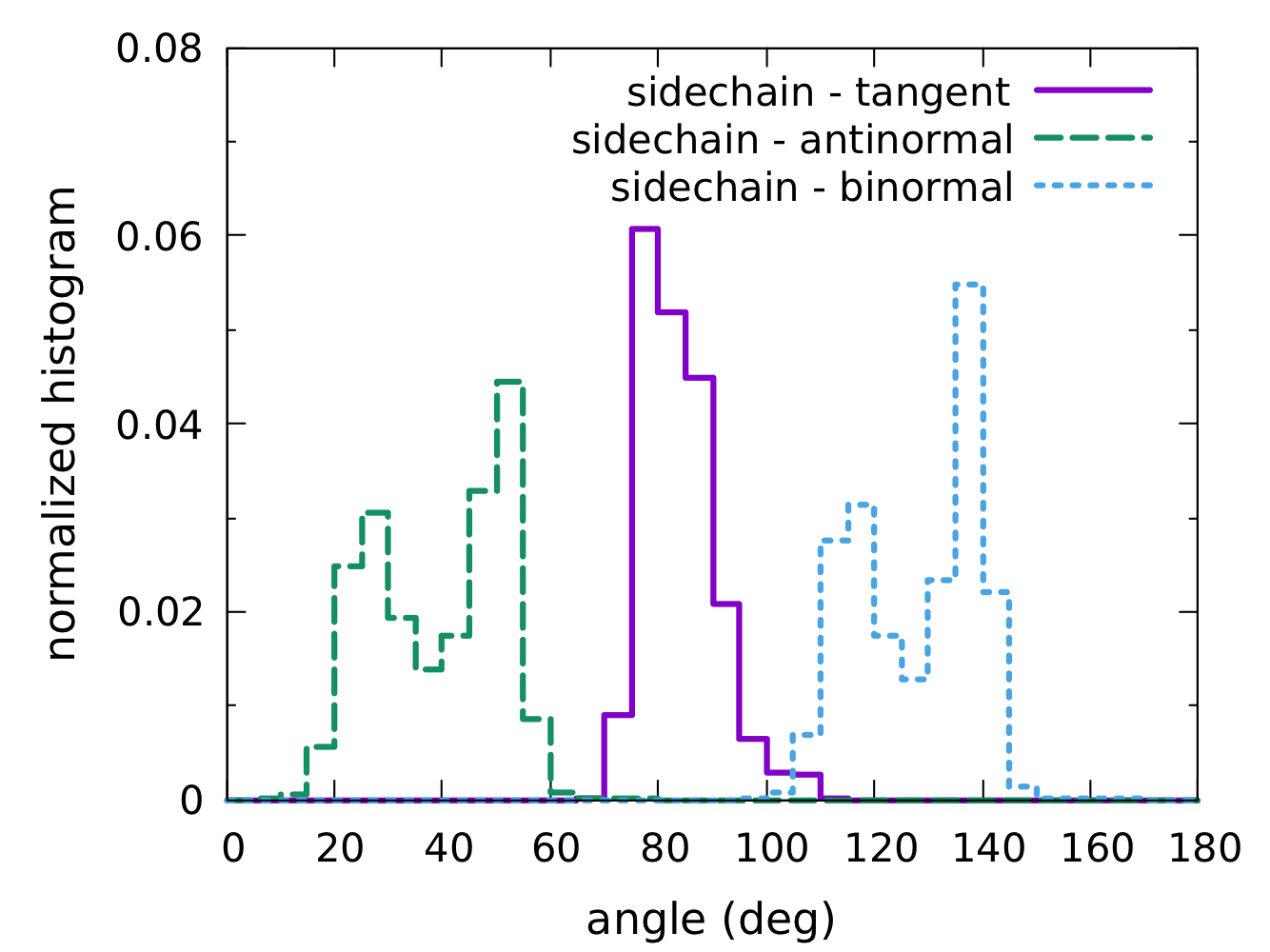}
\caption{Normalized histograms of the tilt angles of amino acids side chain with respect to the tangent (solid), negative normal (dashed) and binormal (dotted) vectors. The data are obtained through the analysis of 500 refined protein structures from the Top500 database. The side chain direction is approximately defined as the C$_\alpha$-C$_\beta$ direction, whereas the vectors of the Frenet frame are calculated based on the coordinates of the C$_\alpha$ atoms. Note that the side chain direction is almost perpendicular to the tangent.}
\label{fig:fig2}
\end{figure}

Denoting the position of the $i$-th C$_{\alpha}$ bead by $\mathbf{r}_i$, the corresponding side chain bead is located at 
\begin{eqnarray}
  \label{sec2:eq1}
\mathbf{r}^{(sc)}_i &=& \mathbf{r}_i - \widehat{\mathbf{N}}_i \frac{\left(\sigma + \sigma_{sc}\right)}{2} ;
\end{eqnarray}
the side chain sphere and the backbone sphere are tangent to each other (see Figure \ref{fig:fig1a}). 
It turns out that the side chains of amino acids in real proteins are roughly oriented along this direction but with an average tilt of roughly $40^{\degree}$ degrees with respect to the negative normal direction (see Figure \ref{fig:fig2}) thus breaking the chiral symmetry, and have different sizes. Our model is greatly simplified -- it does not have any chirality built into it, nor any specificity of the amino acids, but rather aims at accounting for excluded volume effects given by side chains. It is well known from the work by Ramachandran and others \cite{Ramachandran68,Rose06,Rose19b} that steric effects play an important role in the formation of secondary structures. Non-consecutive backbone beads have both steric interactions and short-range attraction of range $R_c$ with all other backbone beads, whereas side chains are subject only to steric interactions.

We will use $\sigma$ as the natural unit of length and study the phase diagram in the three planes of overlap $1-b/\sigma$, size of the side chain $\sigma_{sc}/\sigma$, and attraction range $R_c/\sigma$. A comparison with real proteins can then be carried out by using $\sigma \approx 5 $ {\AA}, the van der Waals sphere of glycine.

\subsection{The Frenet-Serret equations}
\label{subsec:frenet}
In this section, we briefly recall the basic mathematical expressions
from differential geometry and local theory of curves \cite{Coxeter69}, that are routinely used in polymer theory \cite{Kamien02} and will be used to derive the main properties of the thick chain model.
Imagine a tubular object, whose axis is
described by a curve $\mathbf{R}(s)$ parameterized by its 
arc length $0 \le s \le L$. It proves convenient to
introduce a suitable curvilinear coordinate $s$ and a
Frenet frame of unit vectors
$\{\widehat{\mathbf{T}}(s),\widehat{\mathbf{N}}(s),\widehat{\mathbf{B}}(s)\}$
for the tangent, normal and binormal respectively, as follows
\begin{eqnarray}
  \widehat{\mathbf{T}}\left(s\right) &=&
  \frac{\mathbf{R}^{\prime}\left(s\right)}
  {\|\mathbf{R}^{\prime}\left(s\right)\|} \\ \nonumber
\widehat{\mathbf{N}}\left(s\right) &=& 
\frac{\widehat{\mathbf{T}}^{\prime}\left(s\right)}
{\|\widehat{\mathbf{T}}^{\prime}\left(s\right)\|} \\ \nonumber
\widehat{\mathbf{B}}\left(s\right)&=& \widehat{\mathbf{T}}\left(s\right) \times
\widehat{\mathbf{N}}\left(s\right) ,
\label{sec1:eq1}
\end{eqnarray}
where the prime denotes the derivative with respect to the argument. Note that $\|\mathbf{R}^{\prime}\left(s\right)\|=1$ because $s$ is the arc length.
The Frenet coordinates satisfy the Frenet-Serret equations 
\begin{eqnarray}
\frac{\partial \widehat{\mathbf{T}}\left(s\right)}{\partial s} &=& \kappa\left(s\right)
\widehat{\mathbf{N}}\left(s\right) \\ \nonumber
\frac{\partial \widehat{\mathbf{N}}\left(s\right)}{\partial s} &=& - \kappa\left(s\right)
\widehat{\mathbf{T}}\left(s\right) + \tau\left(s\right) \widehat{\mathbf{B}}\left(s\right) \\ \nonumber
\frac{\partial \widehat{\mathbf{B}}\left(s\right)}{\partial s} &=& - \tau\left(s\right)  
 \widehat{\mathbf{N}}\left(s\right) ,
\label{sec1:eq2}
\end{eqnarray} 
which automatically define the curvature $\kappa(s)$ and the torsion $\tau(s)$ from
the first and the last equation. Note that it is conventional to
choose $\kappa(s)$ to be positive by absorbing the sign in the direction of
the normal vector $\widehat{\mathbf{N}}(s)$.
In the simulations, the discrete version of these equations will be exploited
\begin{eqnarray}
  \label{sec1:eq3}
  \widehat{\mathbf{T}}_i&=&\frac{\mathbf{D}_{i}+\mathbf{D}_{i+1}}{\left \vert \mathbf{D}_{i}+\mathbf{D}_{i+1} \right \vert}
\end{eqnarray}
where $\mathbf{D}_i=\mathbf{r}_{i}-\mathbf{r}_{i-1}$ so that $\vert\mathbf{D}\vert=b$. For each of the non-terminal backbone spheres, one defines a normal vector 
\begin{eqnarray}
  \label{sec1:eq4}
\widehat{\mathbf{N}}_i & = & \frac{\mathbf{D}_{i+1}-\mathbf{D}_{i}}
{\vert \mathbf{D}_{i+1}-\mathbf{D}_{i} \vert}
\end{eqnarray} 
where $i=2,\ldots,N-1$. The corresponding binormal vector is then given by
\begin{eqnarray}
  \label{sec:eq5}
\widehat{\mathbf{B}}_i &=& \widehat{\mathbf{T}}_i \times \widehat{\mathbf{N}}_i .
\end{eqnarray}
From this, one can derive the discrete counterparts of Eqs.(\ref{sec1:eq2}) that will automatically define the
discrete curvature $kappa_{i}$ and torsion $\tau_{i}$ (see below).
\subsection{Simulations protocol}
\label{subsec:simulations}
In numerical simulations, we have studied the zero-temperature phase diagram using microcanonical Wang-Landau \cite{Wang01} and conventional simulated annealing canonical Monte Carlo (MC) simulations \cite{Allen89,Frenkel01}, always obtaining consistent results.  In the Wang-Landau simulations, we sample polymer conformations according to the micro-canonical distribution by generating a sequence of chain conformations $A \to B$, and accepting the new configuration $B$ with the micro-canonical acceptance probability
\begin{eqnarray}
\label{sec1:eq19}
P_{acc}(A \rightarrow B)&=&\min{ \left(1,\frac{w_B \, g(E_A)}{w_A \, g(E_B)} \right)},
\end{eqnarray}
\noindent where $w_A$ and $w_B$ are weight factors ensuring the microscopic reversibility of the moves. 
The set of MC moves, that are accepted or rejected according to the probability given by Eq.(\ref{sec1:eq19}), includes both local-type moves, such as single-sphere crankshaft, reptation and end-point, as well as non-local-type moves, such as pivot, bond-bridging and back-bite moves, randomly sampled so that on average $N$ spheres (or a multiple of it) are moved to complete a MC step \cite{Taylor09b}.

The density of states $g(E)$ is then constructed iteratively by filling suitable energy histograms and controlling their flatness. However, in order to compute
the ground state energy, the lowest energy was consecutively selected using the acceptance probability (\ref{sec1:eq19}) with a bias toward less populated energy states. This corresponds to the usual preliminary calculation carried out without a low-energy cut-off in the usual Wang-Landau scheme \cite{Taylor09b}. In the full Wang-Landau calculation, we typically assume convergence after 30 levels of iterations, corresponding to a multiplicative factor value of $f=10^{-9}$.
For the ground state calculation, each run is composed of at least $10^9$ Monte Carlo moves per sphere. In the simulated annealing case, the moves were the same and the temperature was gradually decreased up to reduced temperatures $T^{*}=k_BT/\epsilon=0.01$.
\subsection{Order parameters}
\label{subsec:order}
In Section \ref{sec:results} we will study the low temperature phase diagram of this model that will display the rich polymorphism charactetistic of real proteins ($\alpha$ helix phase, $\beta$ sheet phase, as well as assemblies of  $\alpha$ helices and $\beta$ sheets). Here, different phases will be identified by suitable order parameters. Another way to discriminate between different phases stems from the contact maps that will be introduced further below.

\subsubsection{Torsional order parameter $\tau$}
  Torsion $\tau_{i}$, implicitly included in the discrete counterpart of the Frenet-Serret Eqs. (\ref{sec1:eq2}), whose 
explicit definition can be given in terms of the derivative of $\widehat{\mathbf{T}}_{i}$ as
\begin{eqnarray}
\label{sec2:eq2}
\tau_{i}&=& 
\frac{
\left(\widehat{\mathbf{T}}_{i} \times \widehat{\mathbf{T}}^{(1)}_{i}\right) \cdot 
\widehat{\mathbf{T}}^{(2)}_{i}
}{\left \vert \widehat{\mathbf{T}}^{(1)}_{i} \times \widehat{\mathbf{T}}^{(2)}_{i}\right \vert^2}
\end{eqnarray}
where we have defined $\widehat{\mathbf{T}}^{(n)}_{i}$ as the $n$-th (discrete) derivative of $\widehat{\mathbf{T}}_{i}$. Here, a simple
two (three) points parametrization for $\widehat{\mathbf{T}}^{(1)}_{i}$ ($\widehat{\mathbf{T}}^{(2)}_{i}$) has been used. 

As we will see in Section \ref{sec:results} the probability distribution $p(\tau)$ switches from a unimodal to a bimodal distribution below the coil-helix transition temperature, and hence
$\tau$ will be mainly used to identify the $\alpha$-helix conformation.

\subsubsection{Flatness order parameter}
A key feature of the $\beta$-sheet is to adopt a nearly two-dimensional conformation. Therefore we can distinguish it by computing a flatness order parameter 
  \begin{equation}
    \label{sec2:eq3}
    \langle \widehat{\mathbf{N}}_{i}\cdot (\widehat{\mathbf{N}}_{j} \times \widehat{\mathbf{N}}_{k})\rangle=
    \left\{
    \begin{array}{cc}
      \approx 0 &\qquad \mbox{for a flat phase} \\
      \ne 0 &\qquad \mbox{otherwise}
    \end{array}
    \right.
  \end{equation}
  for all triplets $i,j,k=1,\ldots,N$ of amino acids that are in the $\beta$ phase. A value lower of $\approx 0.2$ of the flatness order parameter will be taken as an indication of the $\beta$ phase.
  
\subsubsection{Radius of gyration parameter}
An important order parameter is given by the mean square radius of gyration $\langle R_g^2(T) \rangle$ that, as in conventional polymers, is able to distinguish between the coil (extended) phase, where the radius of gyration $R^2 \sim N^{2\nu}$, with $2\nu \approx 1.2$ and the globule (collapse) phase, where it is much smaller. In canonical simulations, this is directly accessible, whereas in the Wang-Landau approach it can be obtained as    
\begin{eqnarray}
\label{sec2:eq4}
\left \langle R_g^2(T) \right \rangle &=&\sum_{E} \left \langle R_g^2\right \rangle_{E} g\left(E\right) e^{-E/\left(k_B T\right)}
\end{eqnarray}
where  $\langle R_g^2 \rangle_E$ is the average square of the radius of gyration at fixed energy $E$.

\section{Results}
\label{sec:results}

\subsection{Temperature dependence}
\label{subsec:temperature}
As in conventional polymers, on cooling, here one observes a folding of the chain resembling a second order phase transition (signature of a transition in the case of an infinitely long chain) signalled by a peak in the constant volume heat capacity per monomer $C_V/(N k_B)$ - as usual, the rounding of the peak stems from a finite size effect. Unlike conventional polymers however, where one finds either a direct transition to a crystal or a two-step transition to a globule and then a crystal, depending on the range of attraction $R_c/\sigma$, which is the only controlling parameter \cite{Taylor09b}, here the temperature profile is considerably richer, and depends on the selected values of the parameters.
Each set of the parameters will define a state point in the corresponding phase diagram.
\begin{figure}[htpb]
  \centering
  \captionsetup{justification=raggedright,width=\linewidth}
  \begin{subfigure}{5.4cm}
    \includegraphics[width=\linewidth]{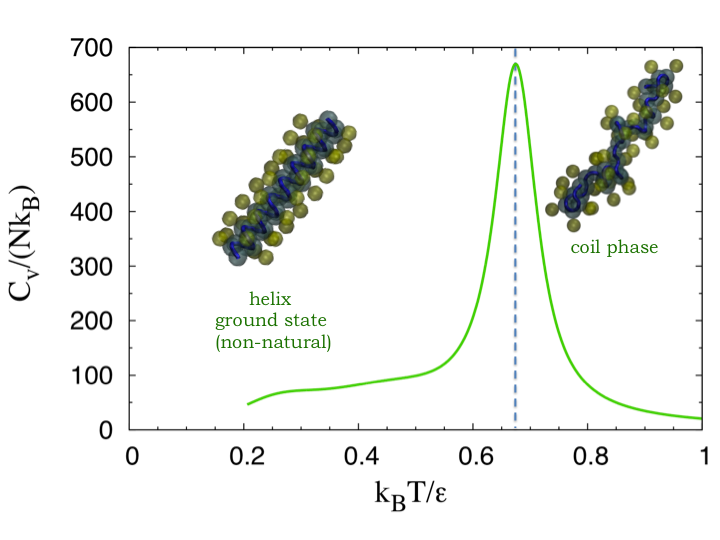}
    \caption{}\label{fig:fig3a}
  \end{subfigure}   
  \begin{subfigure}{5.4cm}
    \includegraphics[width=\linewidth]{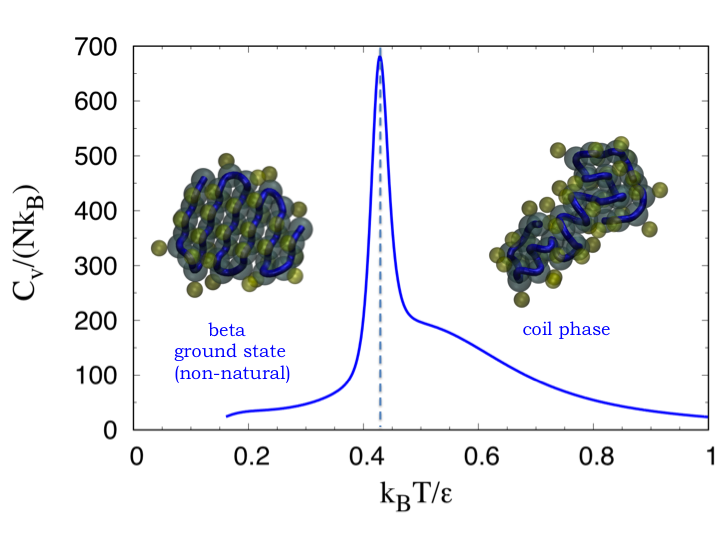}
    \caption{}\label{fig:fig3b}
  \end{subfigure}   
  \begin{subfigure}{5.4cm}
    \includegraphics[width=\linewidth]{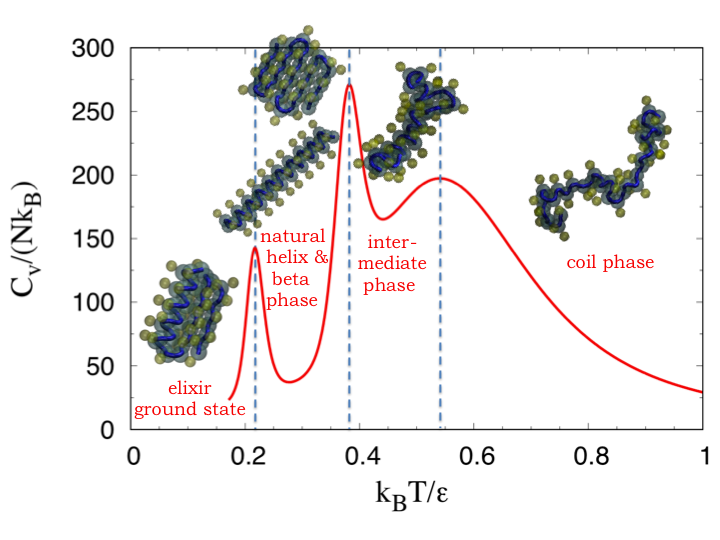}
    \caption{}\label{fig:fig3c}
  \end{subfigure}
    \begin{subfigure}{5.4cm}
    \includegraphics[width=\linewidth]{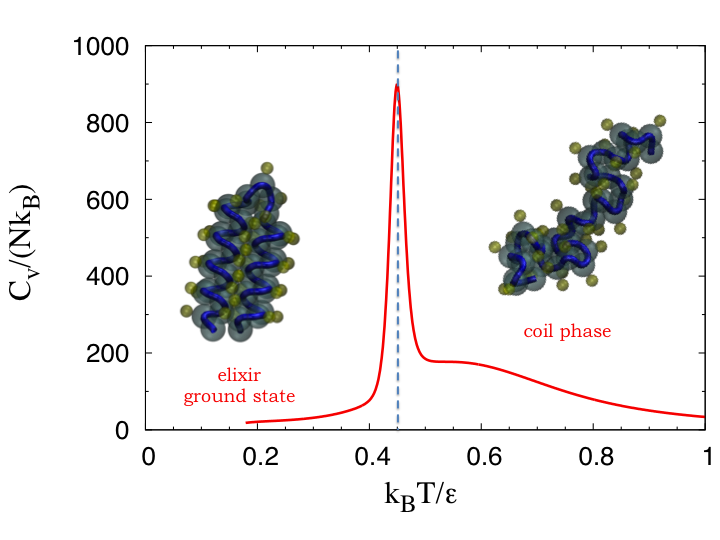}
    \caption{}\label{fig:fig3d}
  \end{subfigure}
	\caption{ (a) The heat capacity per monomer $C_V/(Nk_B)$ as a function of the reduced temperature $k_BT/\epsilon$ for a pathway ending in the helix phase ; (b) Same as in (a) for a pathway ending into the $\beta$ phase; (c) and (d) Same as in (a) for a pathway ending into the degenerate elixir phase. The specific values of the parameters are: (a) $1-b/\sigma=0.25$; $\sigma_{sc}/\sigma=0.83$, $R_c/\sigma=1.17$, transition temperature  $k_B T_{\alpha}/\epsilon=0.67$; (b) $1-b/\sigma=0.18$; $\sigma_{sc}/\sigma=0.67$, $R_c/\sigma=1.167$, transition temperature  $k_B T_{\beta}/\epsilon=0.41$; (c) $1-b/\sigma=0.25$; $\sigma_{sc}/\sigma=0.67$, $R_c/\sigma=1.17$, transition temperature  $k_B T_{elixir}/\epsilon=0.20$; (d) $1-b/\sigma=0.25$; $\sigma_{sc}/\sigma=0.50$, $R_c/\sigma=1.17$, transition temperature  $k_B T_{elixir}/\epsilon=0.45$. In all cases, the chain length is $N=40$.
\label{fig:fig3}}
\end{figure}
For vary large values of the side chain diameter $\sigma_{sc}/\sigma$, the steric effect prevents the collapse of the chain and a coil (swollen) phase is observed at any temperature. In contrast, a high temperature coil state and a low temperature globule phase is observed for low values of  $\sigma_{sc}/\sigma$, separated by a $\Theta$ transition point \cite{deGennes79,Khokhlov02,Rubinstein03}, as in conventional polymers. At intermediate values, however, additional phases are observed depending on the specific value of the parameters. This is the counterpart of what has been denoted as the \textit{marginally compact} phase in the thick chain model that was devised in a spirit similar to the present one \cite{Maritan00}. Here for a large range of parameter values (the nature of the phase diagram will be discussed in depth in the rest of the paper), a single helix phase is found below a characteristic temperature $T_{\alpha}$ that depends on the specific state point. This is shown in Figure \ref{fig:fig3a}, where the transition to a helix state is signalled by a peak in the heat capacity. Note that this is the \textit{true} folded state in this region. A different state point selected in the neighbourhood  still results in the ground state being a single helix with a distinct geometry and a similar transition temperature. Within another wide region of parameter space, a similar heat capacity profile is found, with a single peak in the heat capacity per monomer $C_V/(Nk_B)$ but resulting in a different, almost planar $\beta$-like folded state, as shown in Figure \ref{fig:fig3b}. Again, neighbouring state points have the same $\beta$-like folded ground state (albeit with distinct shapes), so there is a one-to-one correspondence between state points and shapes of the folded states, none of which has a geometry (radius and pitch for the helix states, zig-zag geometry of the strand for the $\beta$-states) comparable to those found in real proteins.

The full low temperature phase diagram as a function of the different parameters will be discussed below. As we will see, there exists a further phase nestled within the above two phases and still within the marginally compact phase. This phase, that will be denoted as \textit{elixir} for reasons that will become clear below, is in fact the most interesting one and was \textit{not} observed in any of the earlier studies, apart from in a recent preliminary report of the present study \cite{Skrbic19,Rose19a}. Here, as shown in Figure \ref{fig:fig3c}, the heat capacity per monomer $C_V/(Nk_B)$ displays multiple peaks upon cooling, first into an intermediate partially folded featureless state, then into a single helix or $\beta$- shape (the helix in the case of Figure \ref{fig:fig3c}), and finally into a superstructure combining both helices and $\beta$s. As in the previous cases, a similar pattern is found for neighbouring state points; unlike the previous cases, however, there is \textit{no} one-to-one correspondence between the state point and final fold, the ground states are now degenerate. As we shall see below, this state point happens to lie close to the boundary of the elixir phase. By considering a state point deeper within the elixir phase (Figure \ref{fig:fig3d}), a higher folding temperature of $T_{elixir}= 0.45 \epsilon/k_B$ is found. This state points will be highlighted in the phase diagrams described below. We note that all these transitions to secondary or assembled secondary structures occur at temperature $T \approx 0.2-0.45 \epsilon/k_B$ that
correspond to attractive energies $\epsilon \approx 2-5$ kcal mol$^{-1}$ at room temperature. Interestingly, this is very close to the accepted value for the strength of a single hydrogen bonds in peptides. \cite{Sheu03}
\subsubsection{Contact Maps}
According to the Levitt-Chothia classification \cite{Levitt76}, all known native states of globular proteins belong to four clearly defined classes:
all-$\alpha$ having only $\alpha$ helix secondary structure, all-$\beta$ having mainly $\beta$ sheets, $\alpha+\beta$ where $\alpha$ helix and
$\beta$ sheet secondary structures do not mix but tend to segregate along the peptide chain, and $\alpha/\beta$ where conversely tend to alternate.
The two latter cases will be generally referred to as $\alpha-\beta$ superstructures.

  Contact maps can be defined by a matrix that is 1 for any two residues that are in contact (i.e. closer than a predefined distance) and 0 otherwise, can also be used as order parameters, both for the all-$\alpha$ and all-$\beta$ phases, and for the combinations of
  $\alpha+\beta$ or $\alpha/\beta$ appearing in the elixir phase. Here, two residues will be considered in contact if
  the distance between the corresponding C$_{\alpha}$s are within the range $R_c$ of the attractive well.

\begin{figure}[htpb]
  \centering
    \captionsetup{justification=raggedright,width=\linewidth}
    \begin{subfigure}{4cm}
      \includegraphics[width=\linewidth]{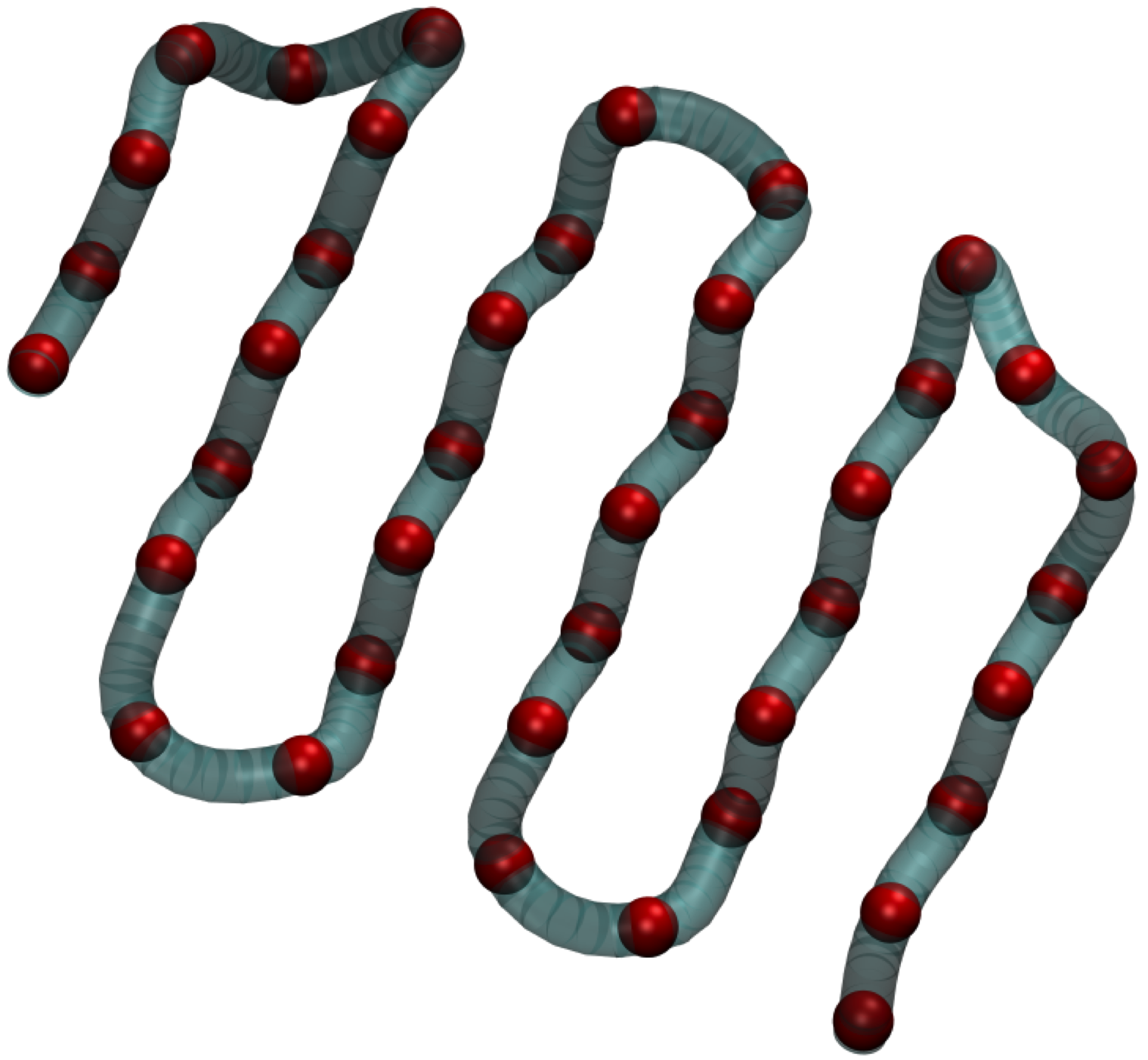}
      \caption{}\label{fig:fig4a}
    \end{subfigure}
     \begin{subfigure}{5cm}
      \includegraphics[width=\linewidth]{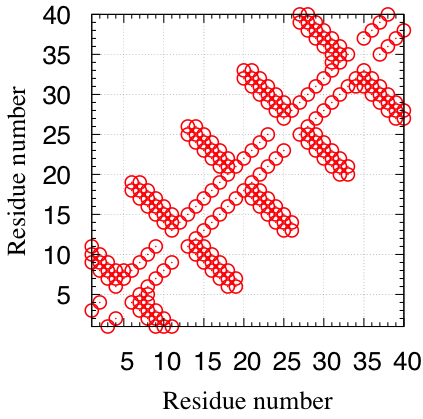}
      \caption{}\label{fig:fig4b}
    \end{subfigure} \\
    \begin{subfigure}{4cm}
      \includegraphics[width=\linewidth]{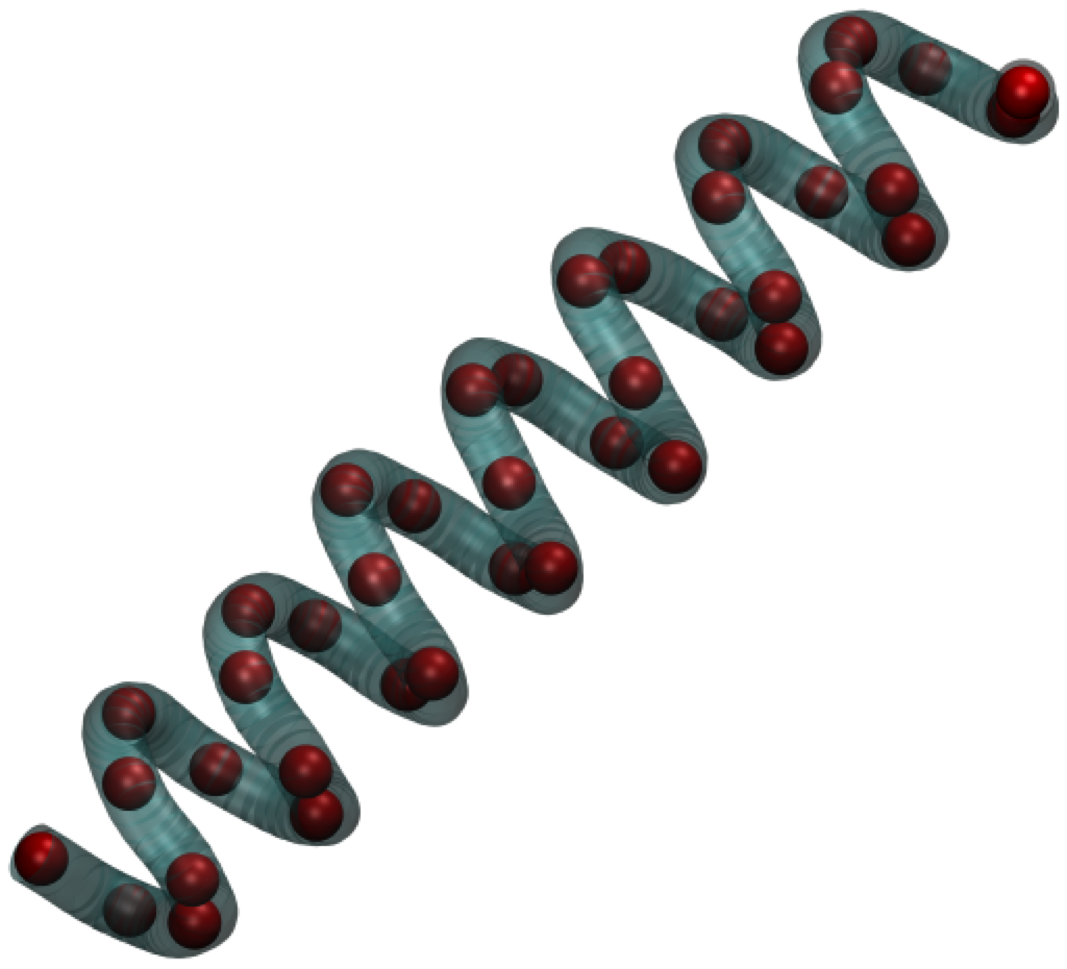}
      \caption{}\label{fig:fig4c}
    \end{subfigure}
    \begin{subfigure}{5cm}
      \includegraphics[width=\linewidth]{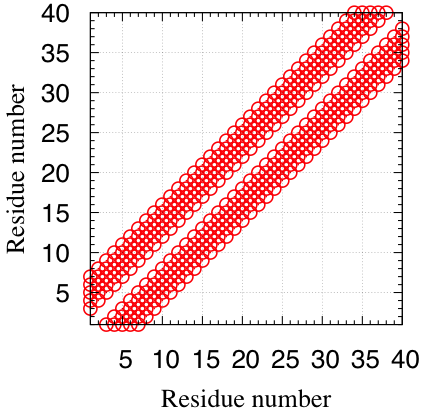}
      \caption{}\label{fig:fig4d}
    \end{subfigure}
  \caption{Results for a $\beta$ sheet structure and an $\alpha$ helix, with $b/\sigma=0.75$ and the side sphere size $\sigma_{sc}/\sigma=0.83$. The figure depicts a snapshot of the ground-state (a,c) and the contact map (b,d) for the cases of $R_c/\sigma=1.1$ (a,b) and $R_c/\sigma=1.4$ (c,d).  A tube representation of the structure, where side chains have been omitted, is shown for clarity. Characteristic fingerprints of the $\beta$ and $\alpha$ structures are visible in the contact maps.
\label{fig:fig4}}
\end{figure}

\begin{figure}[htpb]
  \centering
    \captionsetup{justification=raggedright,width=\linewidth}
    \begin{subfigure}{3.5cm}
     \includegraphics[width=\linewidth]{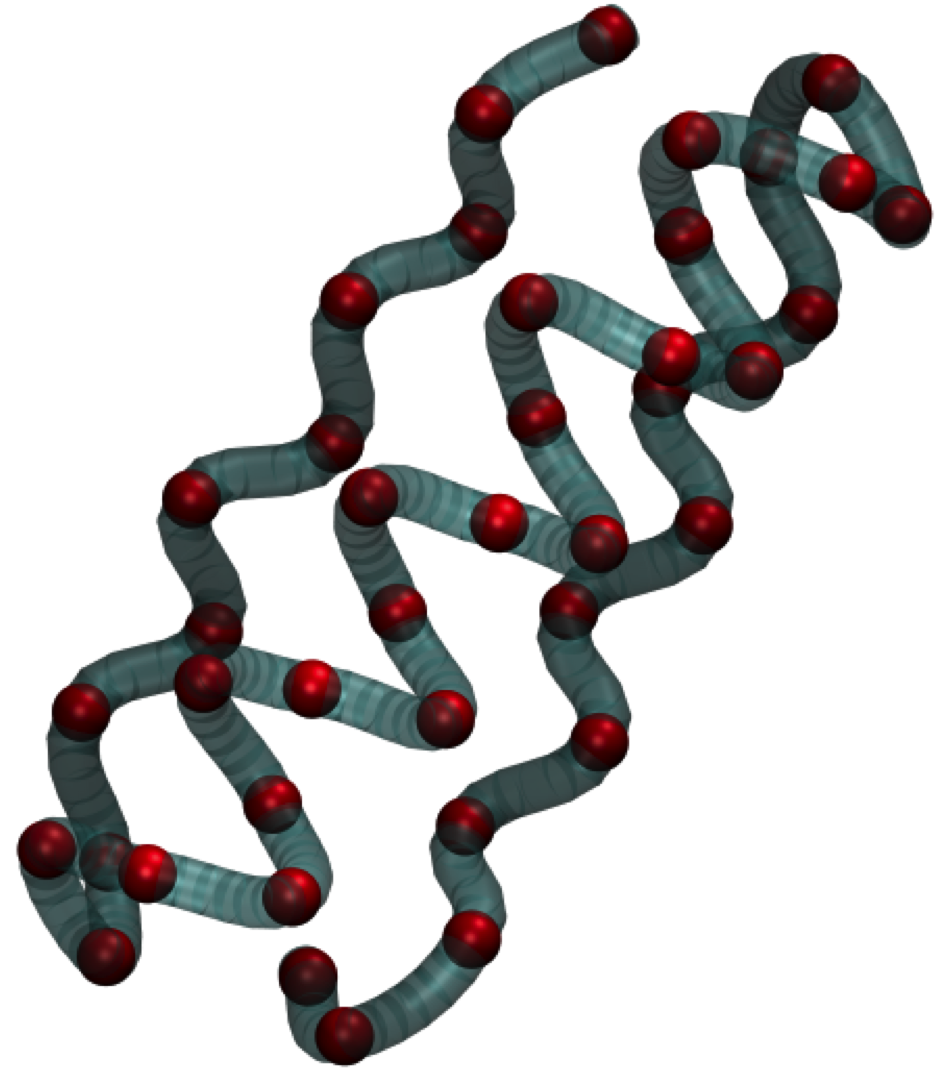}
      \caption{}\label{fig:fig5a}
    \end{subfigure}
     \begin{subfigure}{5cm}
      \includegraphics[width=\linewidth]{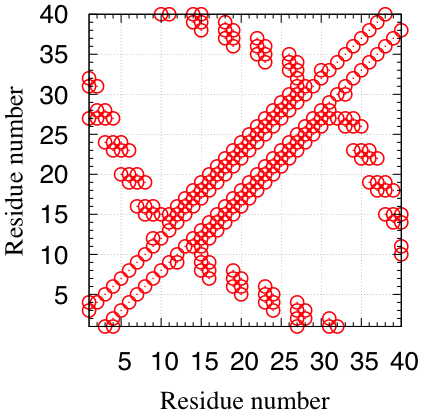}
      \caption{}\label{fig:fig5b}
    \end{subfigure}\\
    \begin{subfigure}{3.2cm}
      \includegraphics[width=\linewidth]{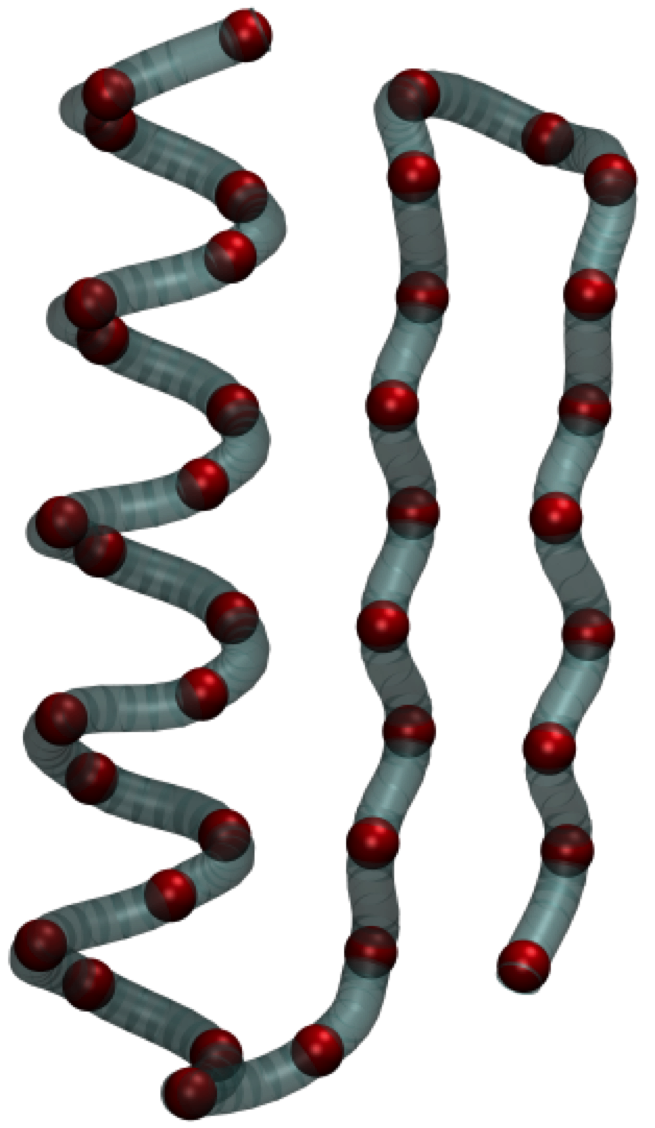}
      \caption{}\label{fig:fig5c}
    \end{subfigure}
    \begin{subfigure}{5cm}
      \includegraphics[width=\linewidth]{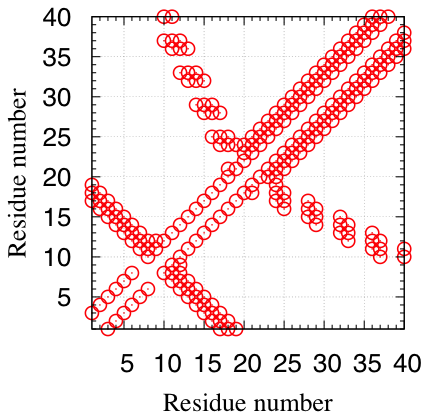}
      \caption{}\label{fig:fig5d}
    \end{subfigure}
  \caption{Representative results in the elixir phase with $b/\sigma=0.75$ and  $R_c/\sigma=1.17$. Displayed are snapshots of the ground-states in tubular representation (a,c) and the contact maps (b,d) for $\sigma_s/\sigma=0.42$ (a,b) and $\sigma_s/\sigma=0.67$ (c,d). Characteristic fingerprints of \textit{both} the $\alpha$ and the $\beta$ conformations within the same structure are clearly visible.
\label{fig:fig5}}
\end{figure}

  Figures \ref{fig:fig4} and \ref{fig:fig5} report the corresponding evidence. 
Figure \ref{fig:fig4b} depicts the contact map associated with the $\beta$-sheet displayed in Figure \ref{fig:fig4a}. The fishbone pattern of the contact map can be unambiguously ascribed to the characteristic shape of the $\beta$-sheet. Likewise, Figure \ref{fig:fig4d} shows a pattern formed by two parallel stripes, that again can be ascribed to the $\alpha$-helix conformation.
Both patterns emerge even in a combined $\alpha-\beta$ superstructure, as illustrated in Figure \ref{fig:fig5}, that shows the contact map and the representative snapshot of a $\alpha/\beta$ (Figures \ref{fig:fig5a} and \ref{fig:fig5b}) and  $\alpha+\beta$ (Figures \ref{fig:fig5c} and \ref{fig:fig5d}). We will find that this is an important characteristic of the elixir phase discussed next.
\subsection{The elixir phase}
\label{subsec:elixir}
\begin{figure}[htpb]
  \centering
    \captionsetup{justification=raggedright,width=\linewidth}
  \begin{subfigure}{5.3cm}
    \includegraphics[width=\linewidth]{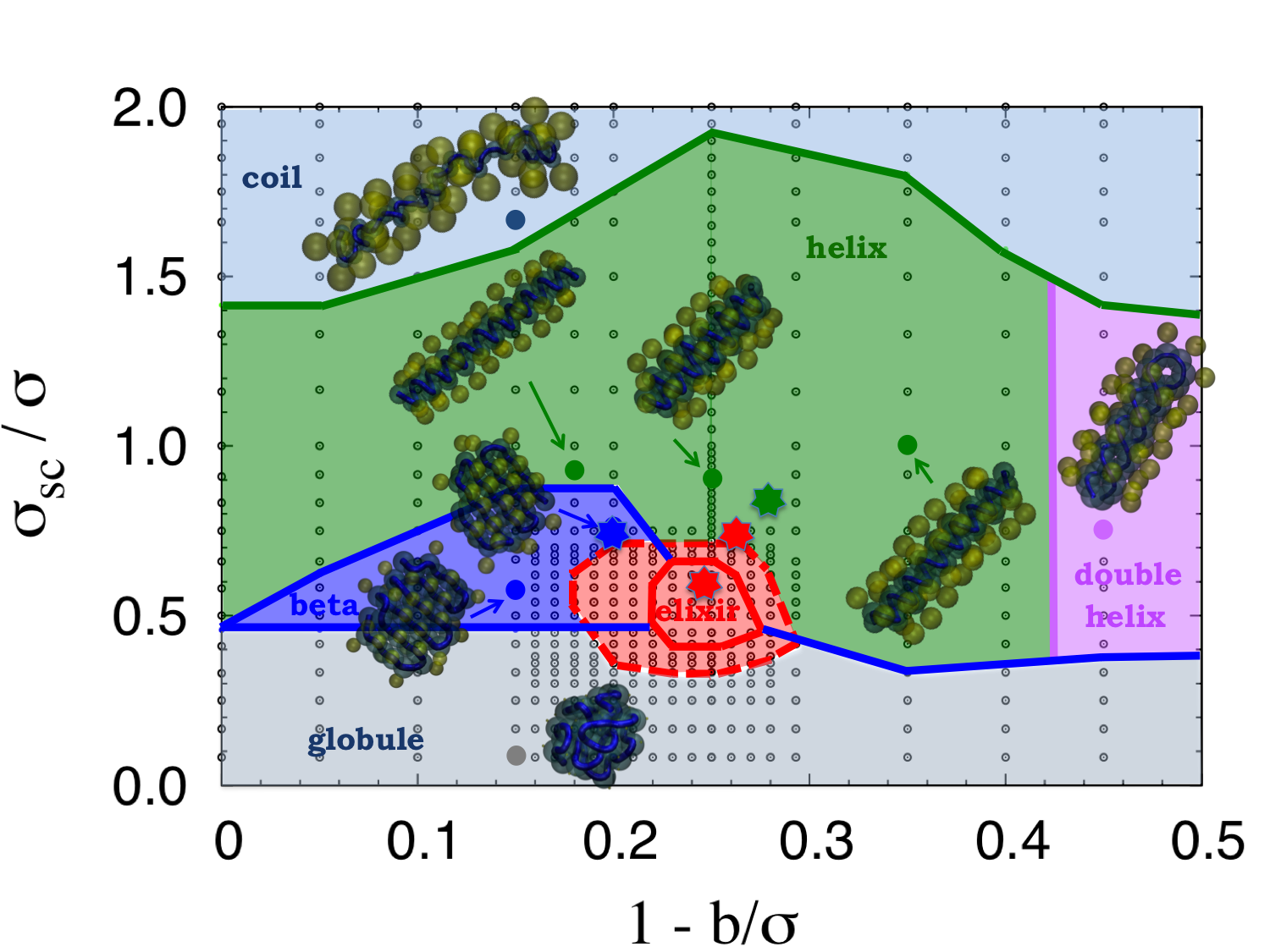}
    \caption{}\label{fig:fig6a}
  \end{subfigure}
  \begin{subfigure}{5.3cm}
    \includegraphics[width=\linewidth]{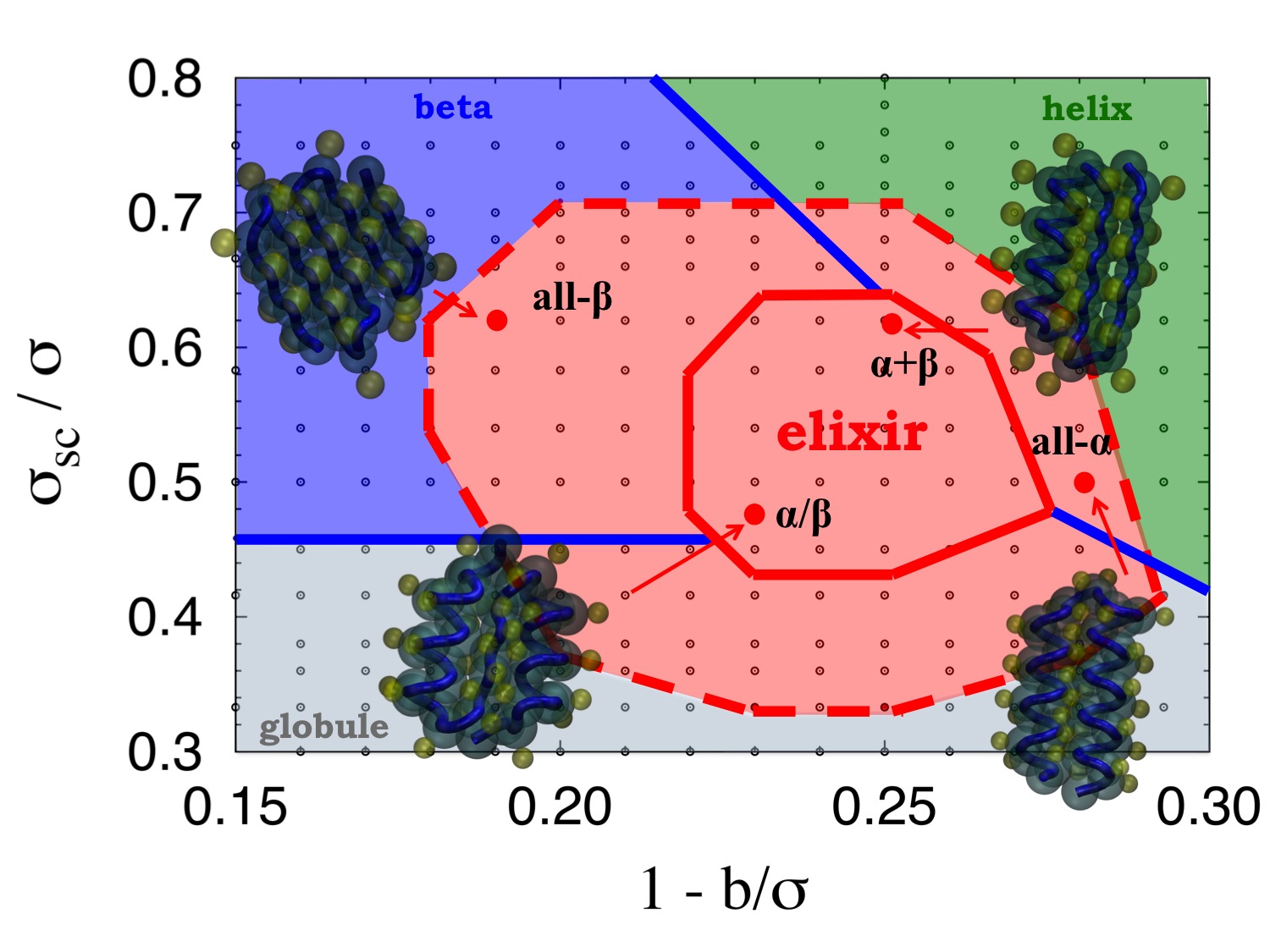}
    \caption{}\label{fig:fig6b}
  \end{subfigure}
    \caption{ (a) The ground state phase diagram along the plane $\sigma_{sc}/\sigma$-$(1-b/\sigma)$; Here $N=40$ and the third non-varying variable has been set to the center of the elixir phase $1-b/\sigma=0.25$; $\sigma_{sc}/\sigma=0.25$; $R_c/\sigma=1.16$. (b) Blow-up of the elixir phase displaying all four prototypical folds of real proteins: all-$\beta$ (close to the $\beta$ and globular phases), all-$\alpha$ (close to the helix and globular phases), $\alpha+\beta$ and $\alpha/\beta$ (inside the inner solid line). Small dots indicate computed state points, and snapshots represent the ground state folds of that particular state point (indicated with a larger dot). In (a), the state points corresponding to the four transitions discussed in Figure \ref{fig:fig3} are highlighted with stars color coded with the corresponding colors.  
\label{fig:fig6}}
\end{figure}
We denote as \textit{ground state} the stable folded state obtained below the folding temperature and study its phase diagram in the space of the three parameters
$1-b/\sigma$, $\sigma_{sc}/\sigma$, and $R_c/\sigma$. Figure \ref{fig:fig6} shows the projection of the phase diagram along the plane $\sigma_{sc}/\sigma$-$(1-b/\sigma)$, with the additional two planes reported in Figure S1 of Supplementary Information (SI). Consider the first $\sigma_{sc}/\sigma$-$(1-b/\sigma)$ plane depicted in Figure \ref{fig:fig6a}, where the marginally compact phase is the whole phase nested between the coil and the globule phases.
Here, two subphases (the helix and the $\beta$) meet with the globular phase within an extended region, denoted as the \textit{elixir phase} \cite{Skrbic19} and delimited by a solid line.
Because different phases meet at this ``extended triple point'', all folds within this region (the elixir phase) must have approximately the same energy. We will assume two folds to have the same energy when their number of contacts do not differ by more than $5\%$.

A similar feature occurs in the other two planes $R_{c}/\sigma$-$(1-b/\sigma)$ and $R_{c}/\sigma$-$\sigma_{sc}/\sigma$ (Figure S1 in SI) where
in all cases there is a well-defined region (the elixir phase) where the helix, $\beta$ and globule phases merge. The difference in shapes of the elixir phase along the three planes can be ascribed to the fact that they are in fact projections of a three dimensional volume along the three different planes. The elixir phase is centered in the state point: $1-b/\sigma=0.25$, $\sigma_{sc}/\sigma=0.5$, $R_c/\sigma=1.167$.  Using $\sigma=5$ {\AA}, the diameter of the van der Waals sphere associated with a Glycine (GLY) residue, one deduces $b=3.81$ {\AA}, $\sigma_{sc}=5$ {\AA}, and $R_c=6$ {\AA}, as previously anticipated.

Additional insights can be obtained by zooming into the elixir phase. This is done in Figure \ref{fig:fig6b} for the $1-b/\sigma$, $\sigma_{sc}/\sigma$ plane of Figure \ref{fig:fig6a}. As discussed, the elixir phase is formed by supersecondary structures of both $\alpha/\beta$ and $\alpha+\beta$ types of many different topologies but nearly identical energies, as shown in Figure \ref{fig:fig6b}. Outside the elixir phase, there is however a larger region (delimited by a dashed line) that
includes also the all-$\alpha$ and all-$\beta$ folds characteristic folds, the remaining two classes in the Levitt-Chothia classification \cite{Levitt76}.  While these have slightly different energies (lower for all-$\beta$, higher for all-$\alpha$ ), all different folds within this larger region have the crucial property of having structural parameters matching those of real proteins. Essentially what happens is the following. On moving from an all-$\beta$ phase into the elixir phase, the structural parameters (length of each single strand, $i$-$i+2$ angle, etc.) gradually changes until they match those found in real proteins upon entering into the elixir phase. Conversely, on moving from an all-$\alpha$ phase into the elixir phase, a morphological transition occurs upon entering the elixir phase driven by an energetic gain in term of the number of contacts. This transition is anticipated by a gradual tuning of the helix structural parameters (radius and pitch) to the correct values matching those of real proteins, within the larger structural basin in the phase diagram delimited by a dotted line. This self-tuning of the single helix allows it to have the number of contacts comparable with those occurring in the $\beta$-phase and hence compete in energy. 

The elixir phase has three remarkable characteristics. First, it includes conformations with nearly identical energies that are composed by a combination of $\alpha$ helices and $\beta$ sheets. Second, all the $\alpha$ helices and $\beta$ sheets found here have geometrical parameters matching those of real proteins. Finally, its spatial extension in parameter space is nearly independent of the number $N$ of backbone beads (i.e. amino acids).
We elaborate more on each of these points in the following Sections.
\subsection{Thermal switching}
\label{subsec:switching}

\begin{figure}[htpb]
  \centering
  \captionsetup{justification=raggedright,width=\linewidth}
    \includegraphics[width=\linewidth]{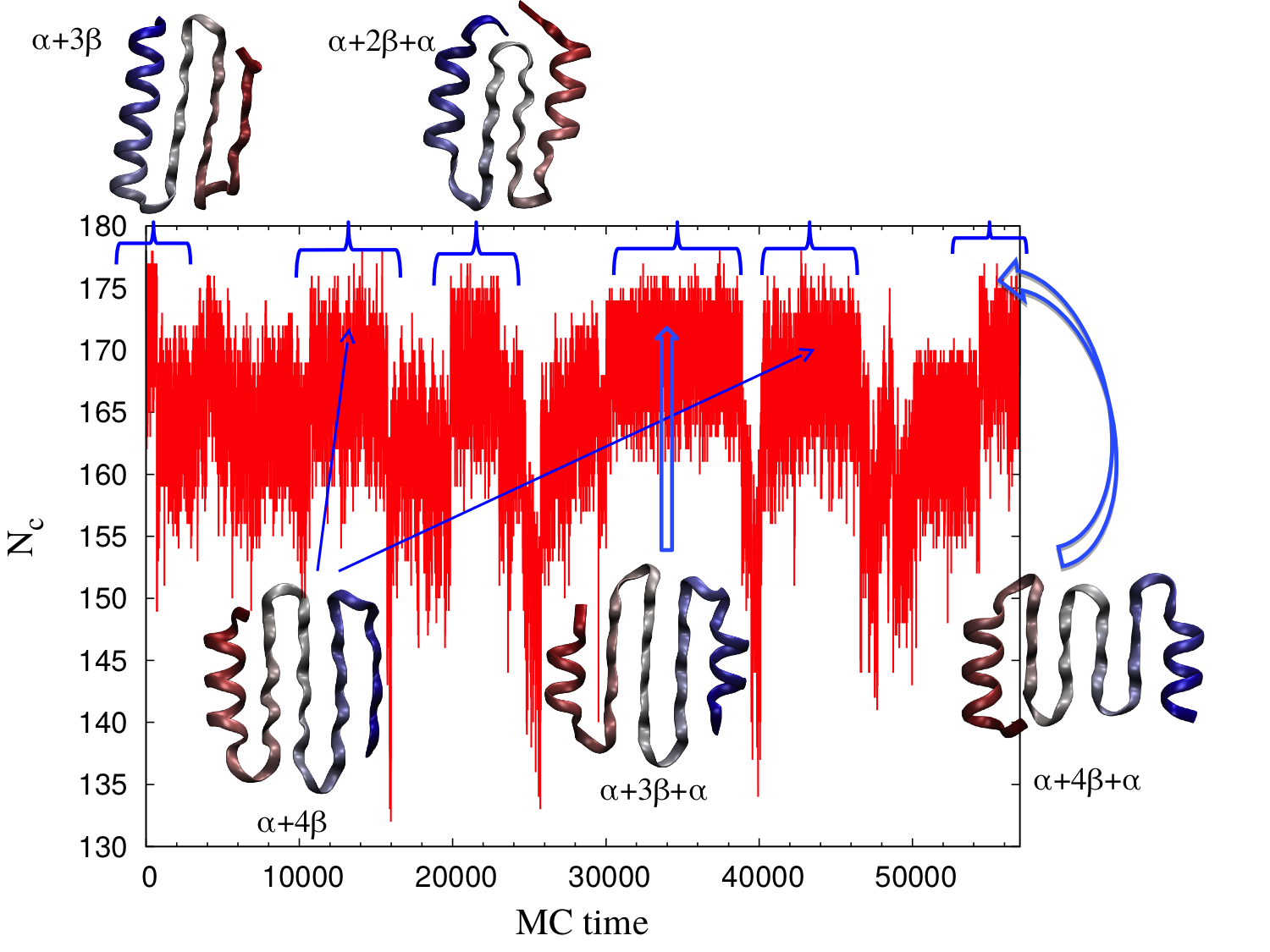}   
  \caption{ Thermal switching at $k_BT/\epsilon=0.35$. Number of contacts $N_c$ as a function of the MC time starting from the initial ground state $\alpha+3\beta$ with $N_c=185$ for chain length $N=56$. Snapshots corresponding to different folds obtained via structural transitions are highlighted.
\label{fig:fig7}}
\end{figure}
The energy degeneracy present in the elixir phase means that at zero temperature (i.e. at temperatures below the folding temperatures $k_BT_{elixir}/\epsilon$  ) it would be possible to go from one such fold to another one due to thermal fluctuations. In order to get further insights on this mechanism, we performed the following \textit{thermal switching} at temperatures slightly lower than folding temperature (Figure \ref{fig:fig7}). We first identified in the central part of the elixir phase 5 different folds of nearly identical energies , each of them obtained upon cooling down from high temperature to the elixir phase below the folding temperature  $k_BT_{elixir}/\epsilon \approx 0.45$. These are depicted in Figure S2 of SI in the case of $N=56$. We then used one of them (the $\alpha+3\beta$ fold having $N_c=185$ contacts) as a starting point of MC calculations carried out at constant temperature $k_BT/\epsilon=0.35$ below the folding temperature. As clearly visible in Figure \ref{fig:fig7}, after an initial sudden drop to a lower number of contacts, the chain starts to probe other possible favourable folds. In the process of doing this, the chain finds other possible folds belonging to the elixir phase, including those other four originally shown in Figure S2 (SI). This switching from one fold to another can be exploited in practical terms to set up nanomachines that rely on conformational changes \cite{Banavar09}.
\subsection{The finite size effects in the elixir phase}
\label{subsec:finite}
\begin{figure}[htpb]
  \centering
    \captionsetup{justification=raggedright,width=\linewidth}
  \begin{subfigure}{7cm}
    \includegraphics[width=\linewidth]{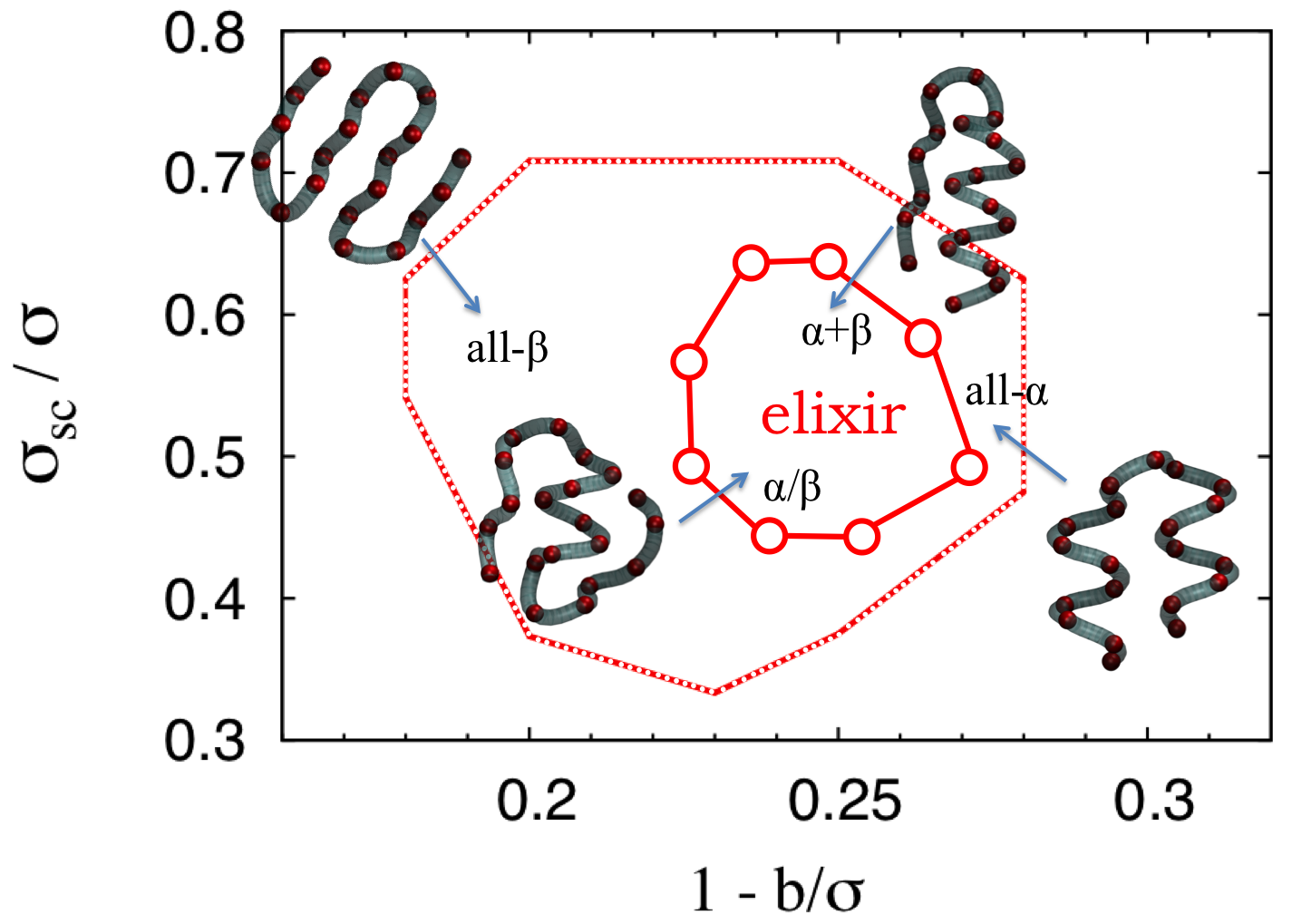}
    \caption{}\label{fig:fig8a}
  \end{subfigure}
    \begin{subfigure}{7cm}
    \includegraphics[width=\linewidth]{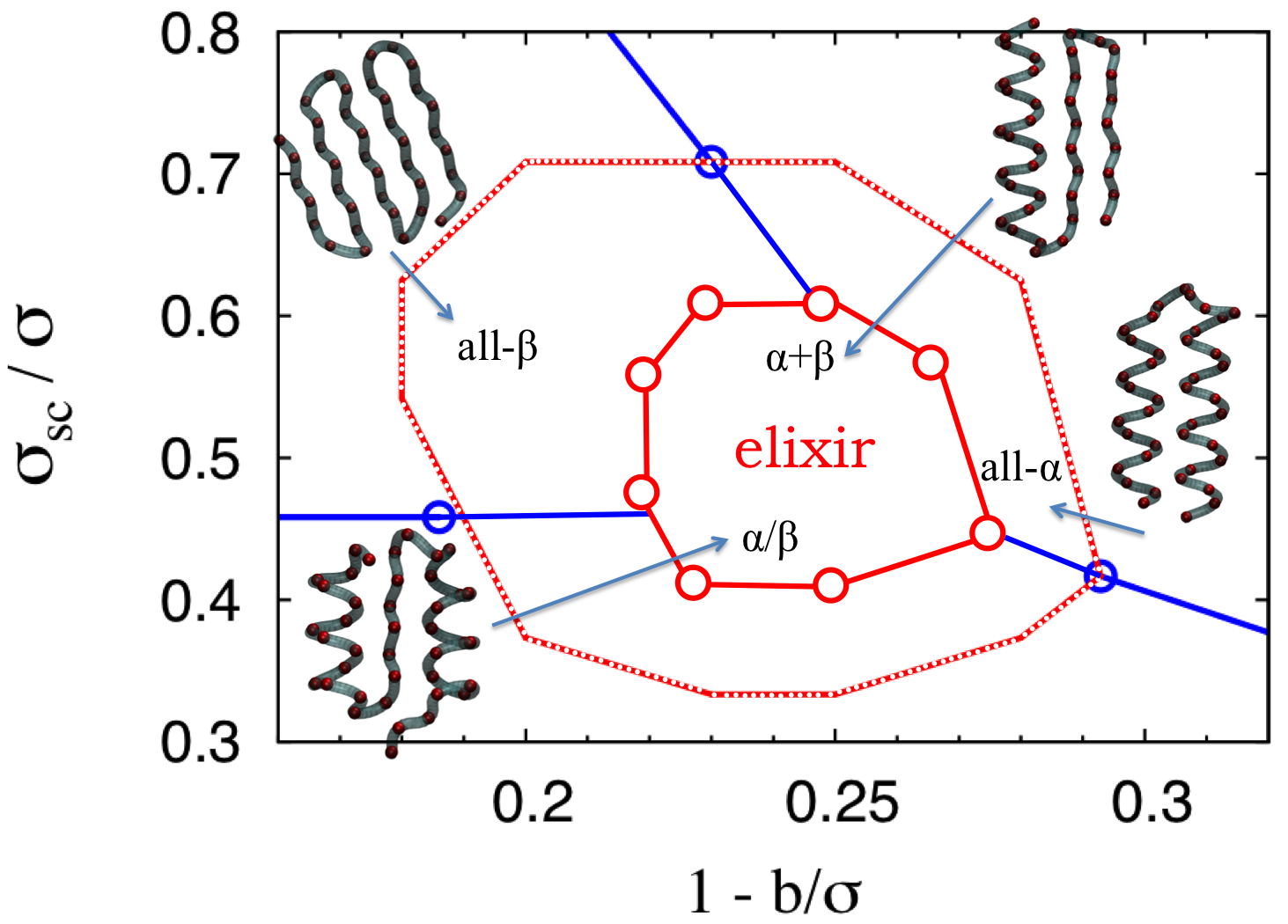}
    \caption{}\label{fig:fig8b}
  \end{subfigure}
   \begin{subfigure}{7cm}
    \includegraphics[width=\linewidth]{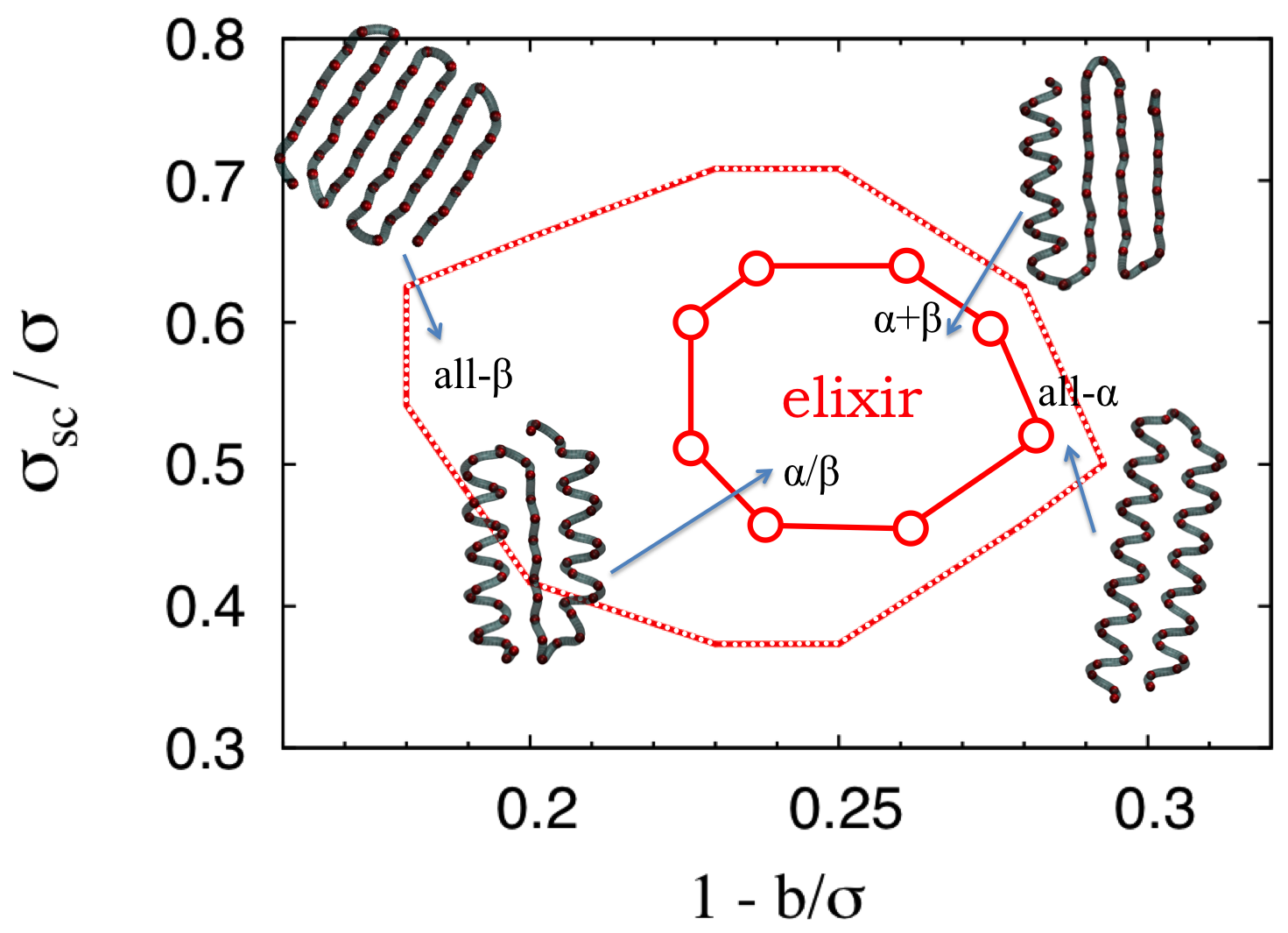}
    \caption{}\label{fig:fig8c}
   \end{subfigure}
  \caption{ The elixir phase  for chain length (a) $N=20$; (b) $N=40$; (c) $N=56$.
\label{fig:fig8}}
\end{figure}

Although most of the results presented here refer to the case $N=40$, we explicitly checked that the size of the elixir phase in the phase diagram and its nature are robust to chain length variations in the range $20 \le N \le 100$. Figure \ref{fig:fig8} shows three representative examples for $N=20$, $N=40$, and $N=56$ in one of the three planes, clearly showing the near independence of the spatial extent in parameter space of the elixir phase on $N$. The case $N=40$ shown in Figure \ref{fig:fig8b} is in fact identical to Figure \ref{fig:fig6b}. Note that all three plots in Figure \ref{fig:fig8} have the same scale, a clear indication that the size and extension of the elixir phase remains rather stable upon changing $N$. Representative snapshots of the corresponding ground state conformation are also displayed as insets. In addition to the (red) solid line indicating the boundary of the elixir phase, where the ground states are combined superstructures $\alpha/\beta$ and $\alpha+\beta$, Figure \ref{fig:fig8} (as well as Figure \ref{fig:fig6}) show a larger region, enclosed by dotted lines, that includes also all-$\alpha$ and all-$\beta$ conformations that are not strictly part of the elixir phase. As we shall see in the next sections, the elixir phase is characterized by a degeneracy in the ground state conformations that reside in it, whereas the larger region is characterized by the fact that each single motif, be it a $\alpha$ helix or a $\beta$ strand, is a unique ground state and has parameters matching those found in real proteins. Clearly, this is always the case for conformations within the elixir phase as it is always contained within the larger dotted region in Fig.\ref{fig:fig8}.

\subsection{The degeneracy of the elixir phase}
\label{subsec:degeneracy}
The elixir phase can be loosely regarded as a non-zero volume in parameter space, whose size does not vary appreciably with chain length, within which there is co-existence of putative ground state structures. In the elixir phase, a combination of $\alpha$ and $\beta$ structures is found. This was already discussed in \cite{Skrbic19} and reiterated more clearly in Figures S3 of SI, reporting the number of contacts $N_c$ (a measure of the energy) as a function of the three parameters of the model: $1-b/\sigma$ (Figure S3a), $\sigma_{sc}/\sigma$ (Figure S3b), and $R_c/\sigma$ (Figure S3c). The elixir phase is marked by two vertical lines inside which $N_c$ is nearly constant.  The fact that it is not exactly constant is due to the combined effects of the discreteness of $N_c$, ultimately related to the use of a square well attractive potential as well as finite size effects. This is shown in Figure S4a of SI reporting $N_c$ as a function of $1-b/\sigma$ for different lengths $N$ of the chain. A further support to this finding is given by Figure S4b (SI) showing the number of contacts per bead $N_c/N$ as $N$ increases. Here we clearly see that the number of contacts per bead of the elixir phase extrapolate to the common value $\approx 4.0$. In Figure S4b we also note the presence of an all-$\alpha$ helix conformation where $N_c/N \to 4.5$ as $1/N \to 0$, indicating that these structures have more ground state contacts (of course for different parameter values) than those in the elixir phase and hence do not belong to it. These structures are however within the dotted region surrounding the elixir phase in the phase diagram of Figures \ref{fig:fig6} and \ref{fig:fig8}, indicating that they have geometries closely matching those in real proteins (see below for a detailed discussion on this point). Likewise, the all-$\beta$ structures have lower ground state energies than those in the elixir phase. It is only in the parameter region corresponding to the elixir phase that, due to the correct matching of the corresponding geometries, they eventually reach the same energy and hence combine into $\alpha/\beta$ or $\alpha+\beta$ structures.

A comparison with a conventional ferromagnetic Ising model can prove instructive. On cooling down from a high temperature disordered phase, larger regions of identically oriented spins emerge, until one of the two symmetrical ground states (spin up or down) is selected. The Ising ground state is then doubly degenerate. The situation is distinct in spin glasses, where the ground state is highly degenerate. The elixir phase has intermediate, finite but significant, degeneracy with no simple analogues in other systems, to the best of our knowledge.

\subsection{Internal structural transitions}
\label{subsec:internal}
\begin{figure}[htpb]
  \centering
    \captionsetup{justification=raggedright,width=\linewidth}
  \begin{subfigure}{3.5cm}
    \includegraphics[width=\linewidth]{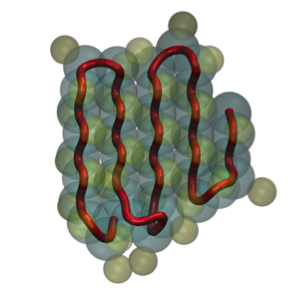}
    \caption{}\label{fig:fig9a}
  \end{subfigure}
    \begin{subfigure}{3.5cm}
    \includegraphics[width=\linewidth]{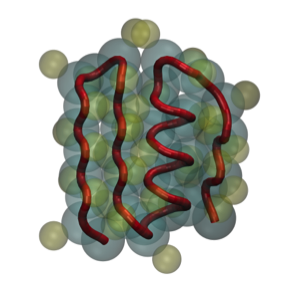}
    \caption{}\label{fig:fig9b}
  \end{subfigure}\\
   \begin{subfigure}{3.5cm}
    \includegraphics[width=\linewidth]{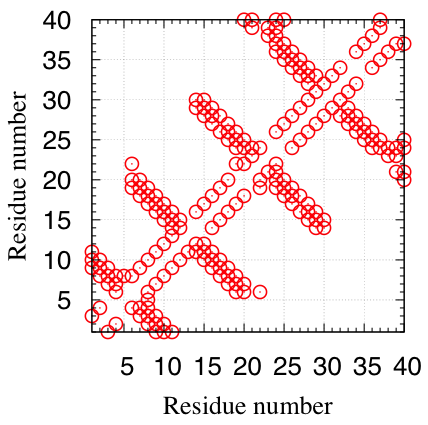}
    \caption{}\label{fig:fig9c}
   \end{subfigure}
    \begin{subfigure}{3.5cm}
    \includegraphics[width=\linewidth]{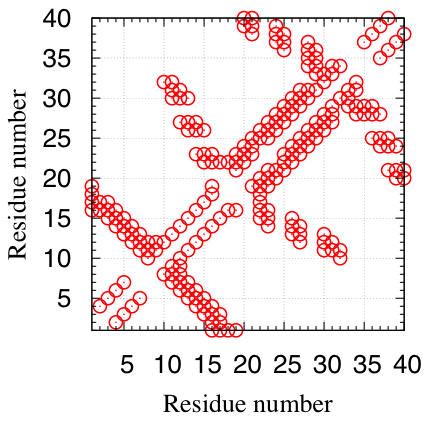}
    \caption{}\label{fig:fig9d}
  \end{subfigure}\\
   \begin{subfigure}{3.9cm}
    \includegraphics[width=\linewidth]{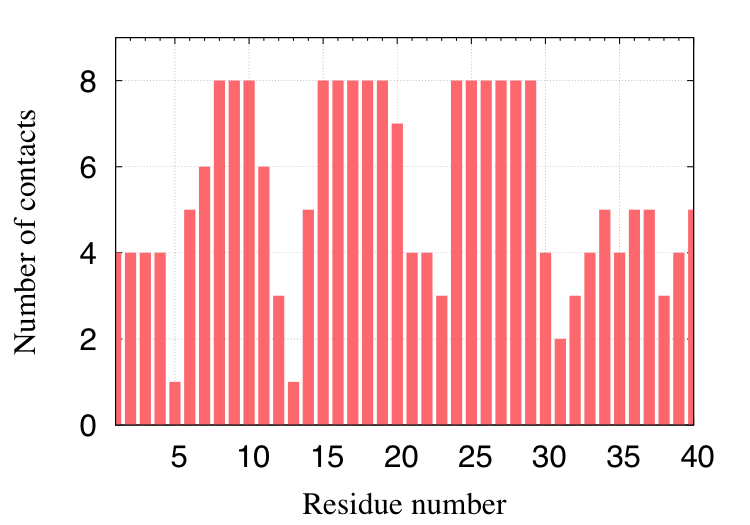}
    \caption{}\label{fig:fig9e}
   \end{subfigure}
   \begin{subfigure}{3.9cm}
    \includegraphics[width=\linewidth]{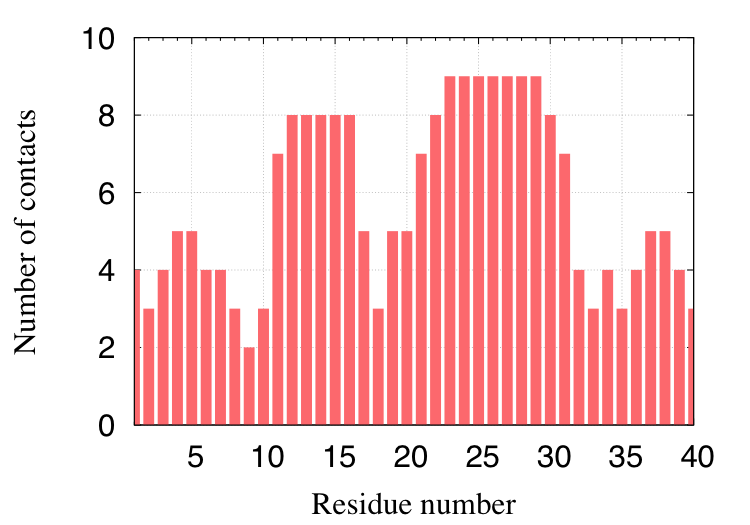}
    \caption{}\label{fig:fig9f}
   \end{subfigure}   
  \caption{Two ground state conformations of a chain in the elixir phase on changing the interaction range of the main chain spheres: a $5 \beta$ structure and a $\alpha+3\beta$ structure. These were obtained for $1-b/\sigma=0.25$, $\sigma_{sc}/\sigma=0.5$  and slightly different ranges of interactions: $R_c/\sigma=1.07$ for the $5 \beta$ structure, and $R_c/\sigma=1.10$ for the $2 \beta+\alpha+\beta$ counterpart. 
    The number of contacts increases from $109$ to $115$ on going from the  $5 \beta$ structure to the $2 \beta+\alpha+\beta$ ground state.(a) Representative snapshot for the $5 \beta$ structure; (b) Representative snapshot for the $2 \beta+\alpha+\beta$ structure; (c) Contact maps for the $5 \beta$ structure; (d) Contact maps for the $2 \beta+\alpha+\beta$ structure; (e) Distribution of contacts among residues for the $5 \beta$ structure; (f) Distribution of contacts among residues for the $2 \beta+\alpha+\beta$ structure;
\label{fig:fig9}}
\end{figure}

\begin{figure}[htpb]
  \centering
  \captionsetup{justification=raggedright,width=\linewidth}
  \begin{subfigure}{3.5cm}
    \includegraphics[width=\linewidth]{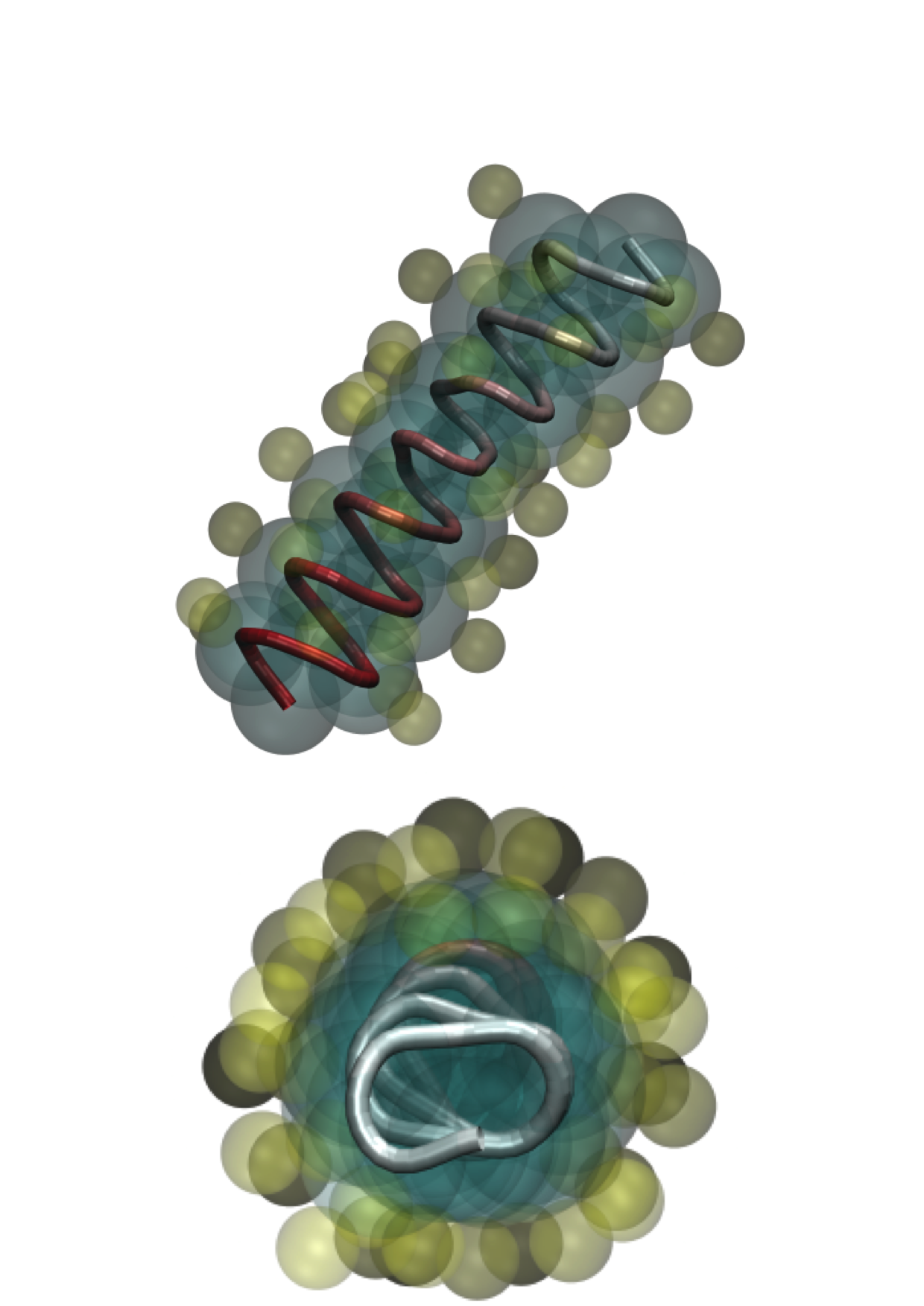}
    \caption{}\label{fig:fig10a}
  \end{subfigure}
   \begin{subfigure}{3.5cm}
    \includegraphics[width=\linewidth]{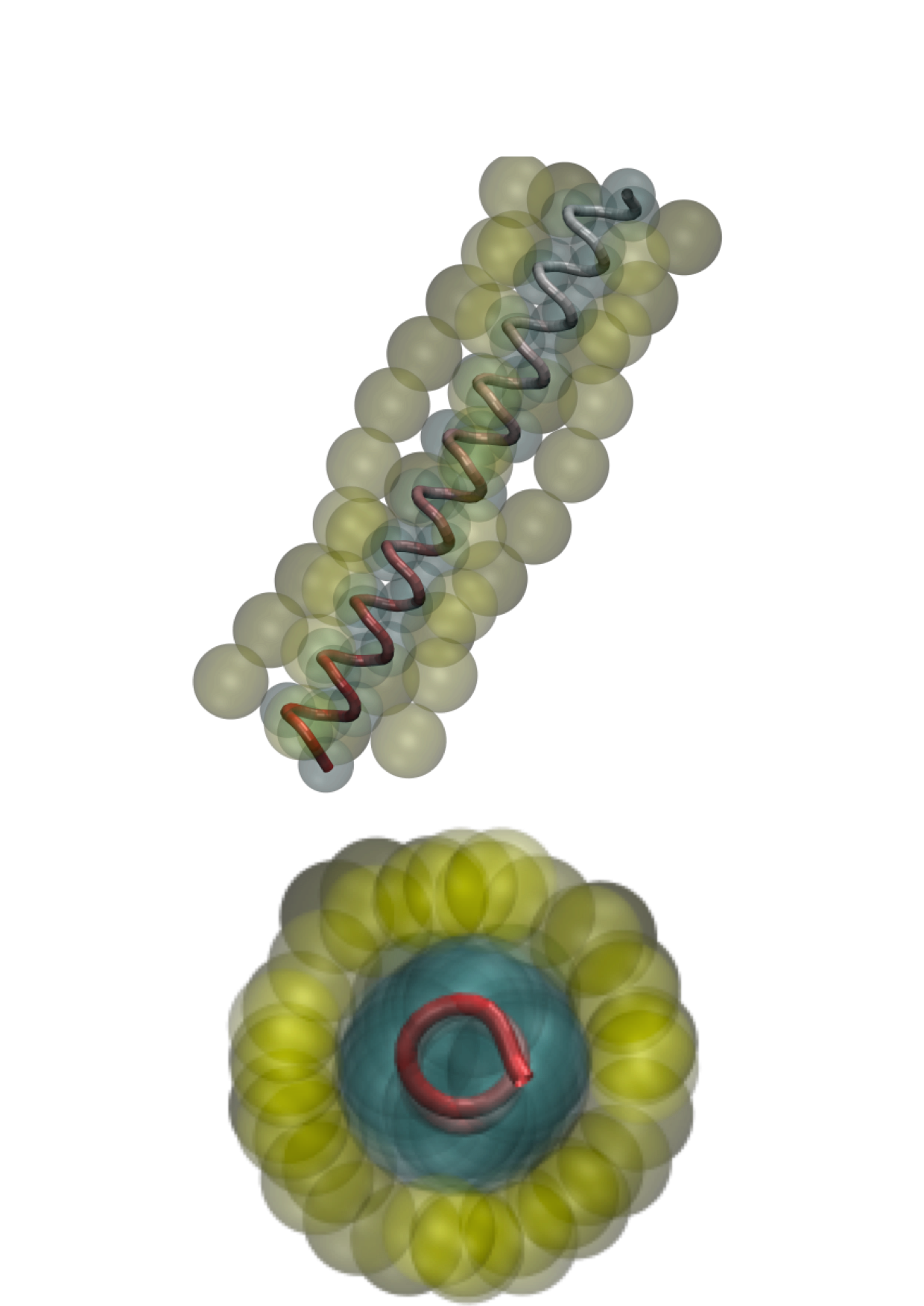}
    \caption{}\label{fig:fig10b}
  \end{subfigure}\\
    \begin{subfigure}{3.5cm}
    \includegraphics[width=\linewidth]{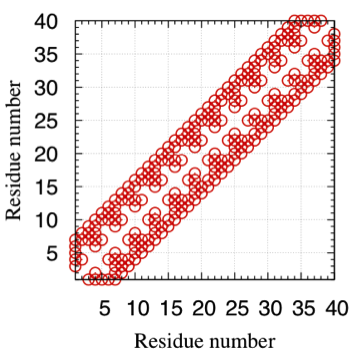}
    \caption{}\label{fig:fig10c}
  \end{subfigure}
   \begin{subfigure}{3.5cm}
    \includegraphics[width=\linewidth]{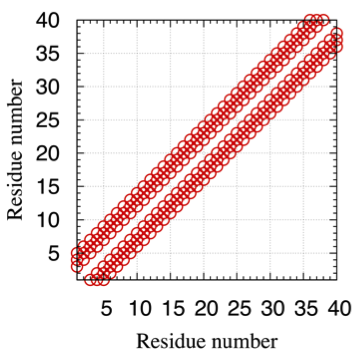}
    \caption{}\label{fig:fig10d}
   \end{subfigure}\\
    \begin{subfigure}{3.9cm}
    \includegraphics[width=\linewidth]{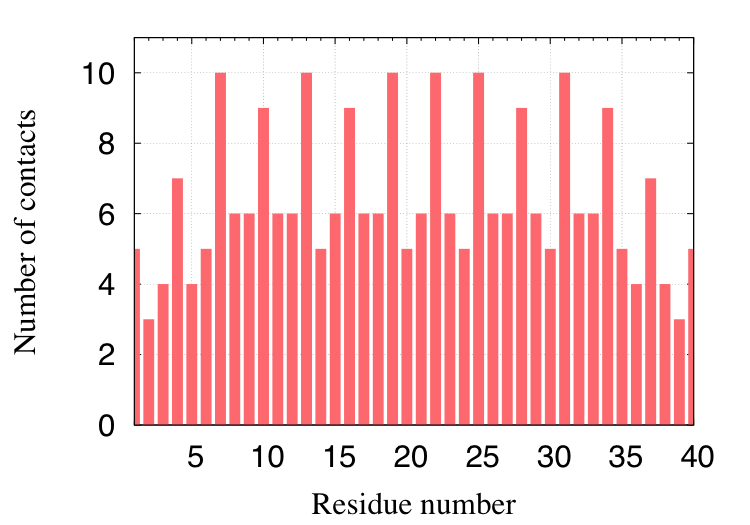}
    \caption{}\label{fig:fig10e}
  \end{subfigure}
   \begin{subfigure}{3.9cm}
    \includegraphics[width=\linewidth]{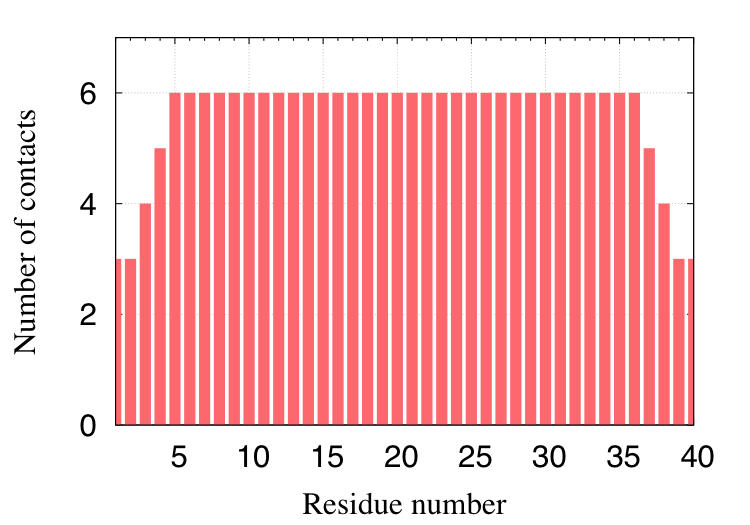}
    \caption{}\label{fig:fig10f}
   \end{subfigure}\\   
   \begin{subfigure}{3.5cm}
    \includegraphics[width=\linewidth]{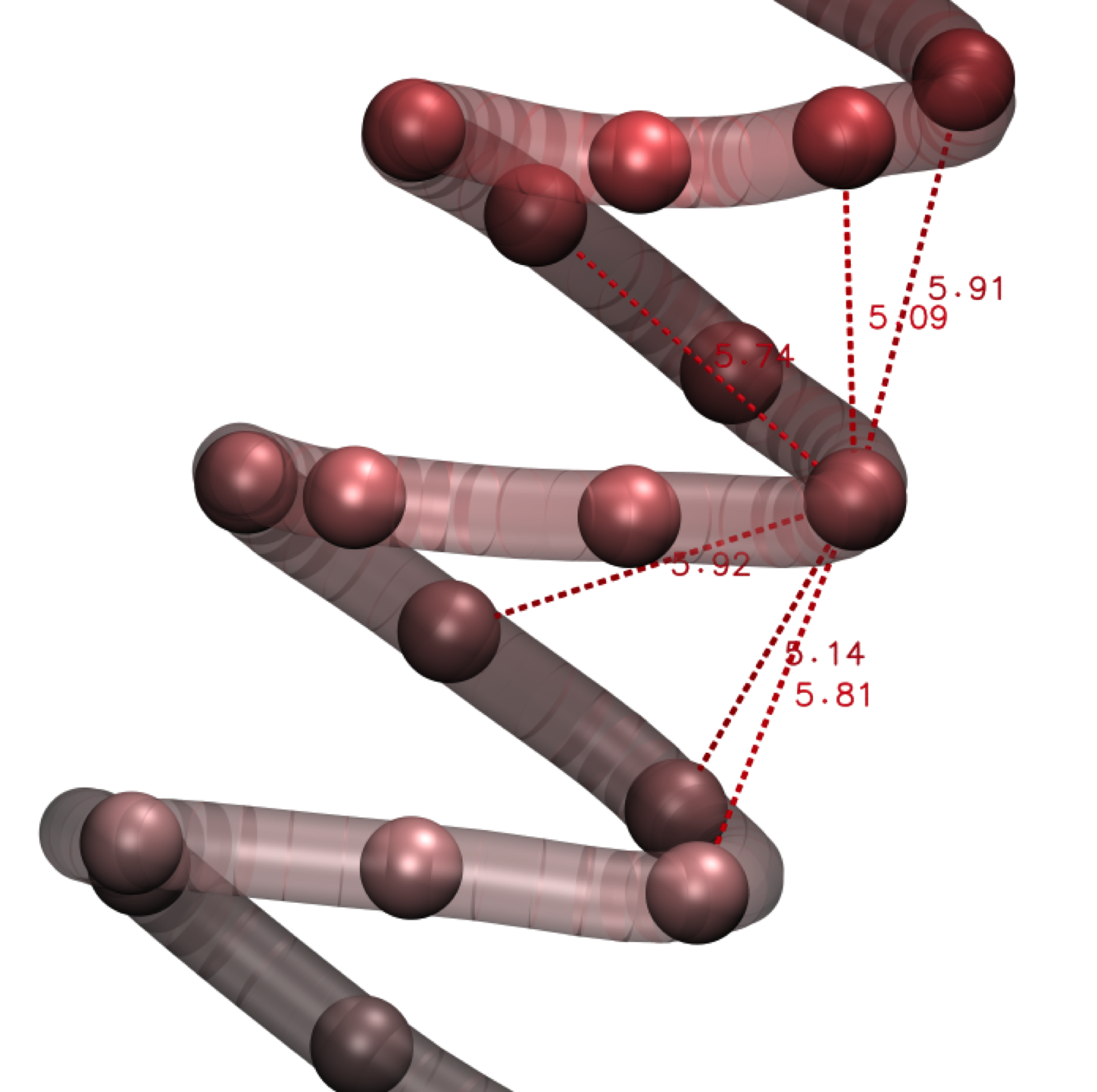}
    \caption{}\label{fig:fig10g}
  \end{subfigure}
   \begin{subfigure}{3.5cm}
    \includegraphics[width=\linewidth]{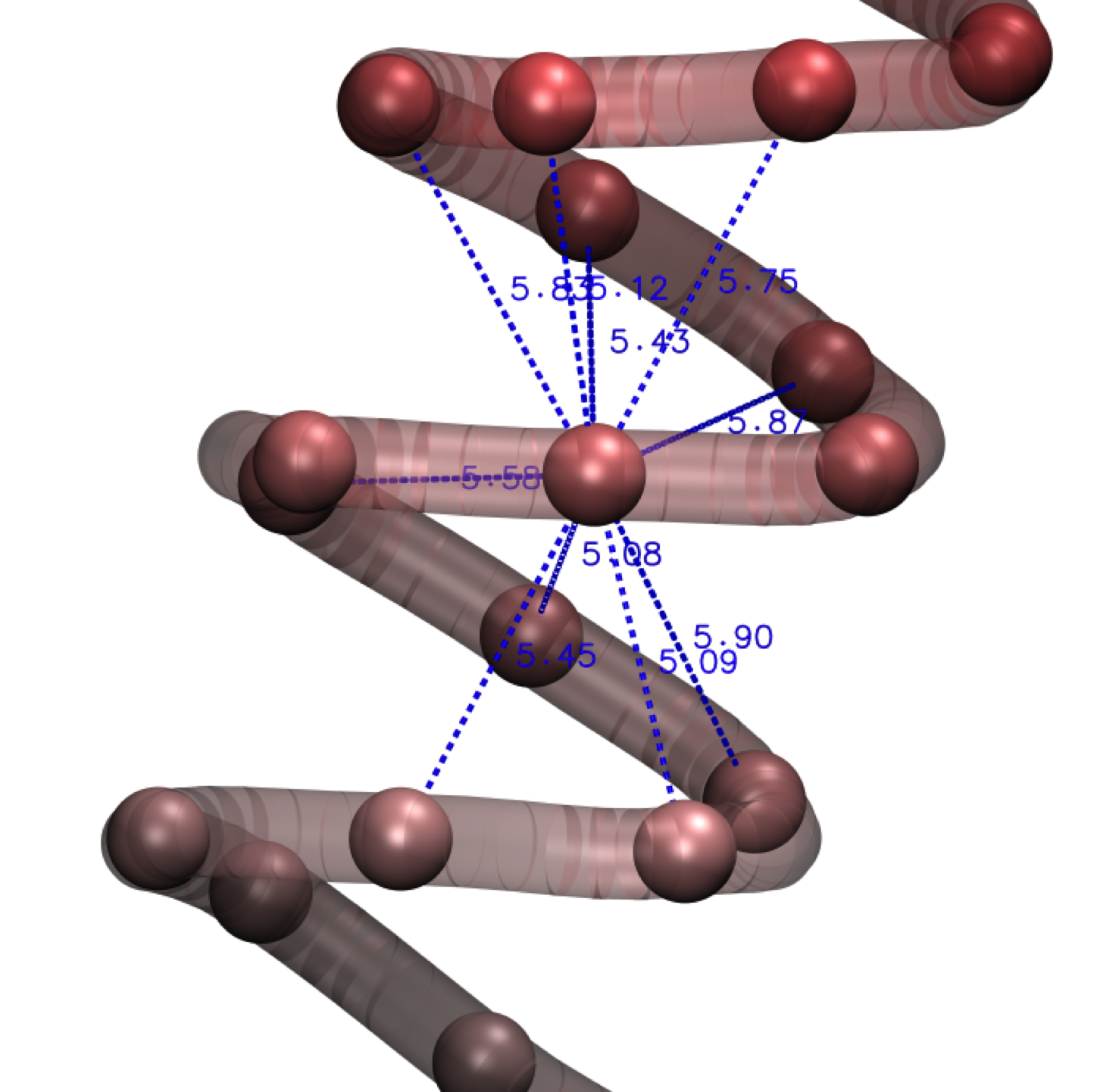}
    \caption{}\label{fig:fig10h}
   \end{subfigure}
  \caption{The helixI$\to$helixII structural transition that is displayed in Figure S3b of SI. Helix I has a higher ($\approx 128$) number of contacts than helix II  ($\approx 111$) because of the smaller sizes of the side spheres thus allowing for higher coordination. Representative snapshots helixI (a) and helixII (b); Contact maps helixI (c) and helixII (d); Distribution of contacts among residues helix I (e) and helixII (f). The last two bottom panels show how the helix I can achieve 6 (g) or 10 (h) contacts.
\label{fig:fig10}}
\end{figure}

The degeneracy of the elixir phase raises the question of how a combination of $\alpha$ and $\beta$ structures can favorably compare with all-$\alpha$ and all-$\beta$ conformations. We address this issue in Figure \ref{fig:fig9} where we monitor the transition from an all-$\beta$ ground state to
a combination of $\alpha+\beta$ conformations.
Figure \ref{fig:fig9} shows how a chain changes its structure from a $5 \beta$ chain ground state to a $2\beta+\alpha+\beta$ upon changing the interaction range driving the system from outside to inside the elixir phase. In real proteins, this can be realized by changing the amino acid sequence. Representative snapshots are depicted in Figures \ref{fig:fig9a} and \ref{fig:fig9a}. This transition is clearly visible on comparing the contact maps of the two conformations, as shown in Figures \ref{fig:fig9c} and \ref{fig:fig9d}. Essentially, the difference is tantamount to a transformation of a $2\beta$ section into an $\alpha$ section. This increases the number of contacts by $6$ units and hence is energetically favourable. The gain is clearly visible in the distribution of the number of contacts $N_c$ among the 40 residues, that shows an increase from 8 to 10 contacts in the region surrounding the 25-th residue (see Figures \ref{fig:fig9e} and \ref{fig:fig9f}).
Another interesting example of a structural change occurs in the helix region in Figure S3b (SI), leading to a step-like discontinuity in the number of contacts, while still preserving the (non-natural) helical shape.
This is addressed in Figure \ref{fig:fig10} that compares the different characteristics of the two resulting helices. Note that they are both outside the elixir phase, so their geometrical parameters are not natural (i.e. not matching those in real proteins). The helix in Figure \ref{fig:fig10a} (denoted as helix I) has a shape that vaguely resembles that of the $\beta$-helix, and has 128 contacts, and it is found just after the elixir phase, as shown in Figure S3b of SI. The helix in Figure \ref{fig:fig10b} (denoted as helix II in Figure S3b of SI), has 111 contacts, and is found for larger values of $\sigma_{sc}/\sigma$. From the contact maps of Figures \ref{fig:fig10c} and \ref{fig:fig10d} we see that helix I is able to achieve a larger number of contacts by switching periodically the number of contacts from 6 to 10 as a function of residue number, at variance with helix II where the bulky side chain prevents this and the number of contacts is 6 in the interior (see Figure \ref{fig:fig10e} and \ref{fig:fig10f}). Figures \ref{fig:fig10g} and \ref{fig:fig10h} show in practice how this is achieved in helix I, with internal beads having 10 contacts and external beads having 6.     
\subsection{Characteristic geometries in real proteins}
\label{subsec:characteristic}
As previously noted, the structures found in the elixir phase display a remarkable similarity with those found in real proteins. The aim of the present Section is to illustrate this point by comparing each of the structural units (helix and $\beta$-strands) of the elixir phase with those of real proteins that are all identical, irrespective of the amino acid specificity of any given protein.

\begin{figure}[htpb]
  \centering
  \captionsetup{justification=raggedright,width=\linewidth}
\includegraphics[width=8cm]{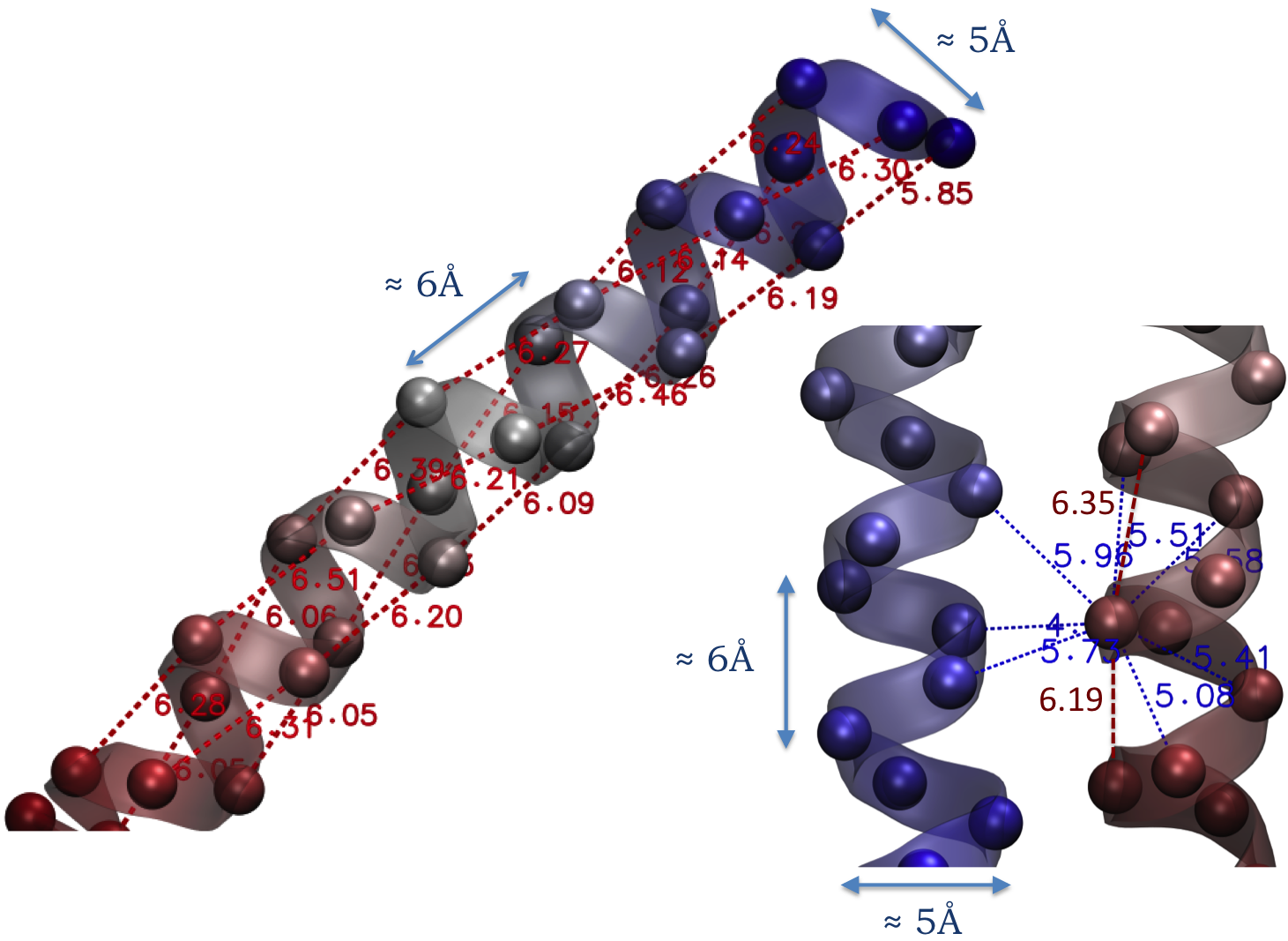}
\caption{Example of all-$\alpha$ structures in real proteins.(Left) One of the $\alpha$-helices of the all-$\alpha$ 1ROP protein.
C$_{\alpha}^i$-C$_{\alpha}^{i+4}$ distances (in {\AA}) of the hydrogen
bonded amino-acids along the helix are highlighted in red. The characteristic lengths
of an $\alpha$-helix are: distance between the neighbouring turns ($\approx$ 6 {\AA}) and its diameter ($\approx$ 5 {\AA}). 
(Right) Association of the two $\alpha$-helices of 1ROP protein:  distances (in {\AA}) of the neighbouring amino-acids ($\approx$ 5-6 {\AA}) from the reference amino-acid (located in the helix on the right); the distances from all other
amino-acids are $\geq 8$ {\AA}. Note that here the highlighted contacts include both two hydrogen bonds within the same helix (in red) and hydrophobic contacts within $\approx$ 6 {\AA} from the reference amino acid (in blue). Six of the contacts  shown are within the same $\alpha$-helix, whereas the remaining three are  formed with amino-acids belonging to the adjacent $\alpha$-helix. In total, this amounts to a typical coordination number of $n_{coord}=9$ for a backbone sphere inside an all-$\alpha$ environment. 
\label{fig:fig11}}
\end{figure}

\begin{figure}[htpb]
  \centering
  \captionsetup{justification=raggedright,width=\linewidth}
\includegraphics[width=8cm]{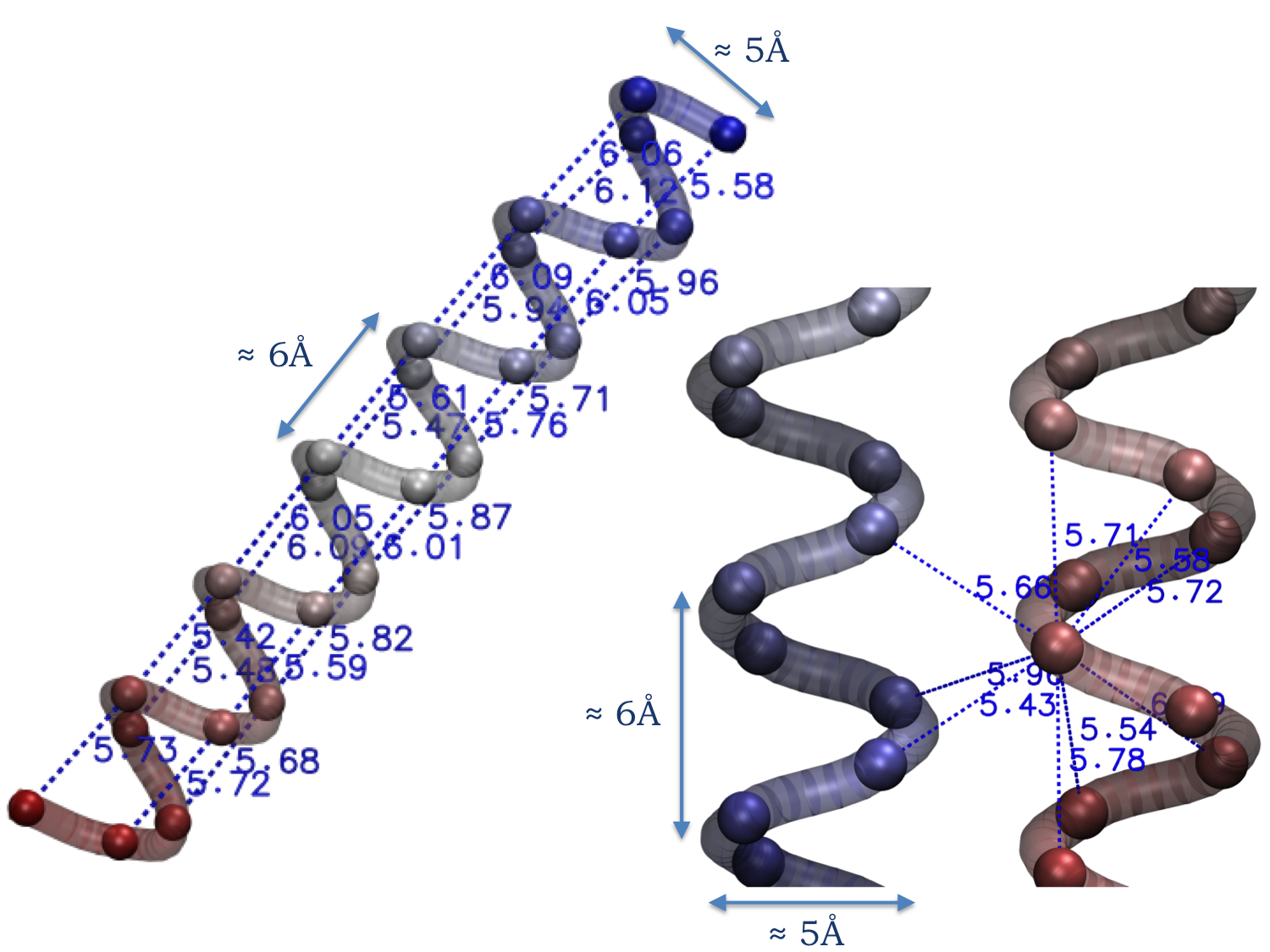}
\caption{Example of all-$\alpha$ structure in the elixir phase. (Left) One of the $\alpha$-helices where each $i$-th residue is within
  $\approx 6 $ {\AA} from the $i-4$ and $i+4$ neighboring residues (highlighted in blue) in accord with the correct geometry of the helix. (Right) Association of two $\alpha$ helices, where there are $6$ contacts within each helix and $3$ additional contacts with the adjacent parallel helix, as in  proteins (see Figure \ref{fig:fig11}). Helices have structural parameters (radii and pitches) such that the average coordination number for a sphere embedded in all-$\alpha$ environment is $n_{coord}=9$, as in  proteins.
  \label{fig:fig12}}
\end{figure}

\begin{figure}[htpb]
  \centering
  \captionsetup{justification=raggedright,width=\linewidth}
  \begin{subfigure}{5cm}
    \includegraphics[width=\linewidth]{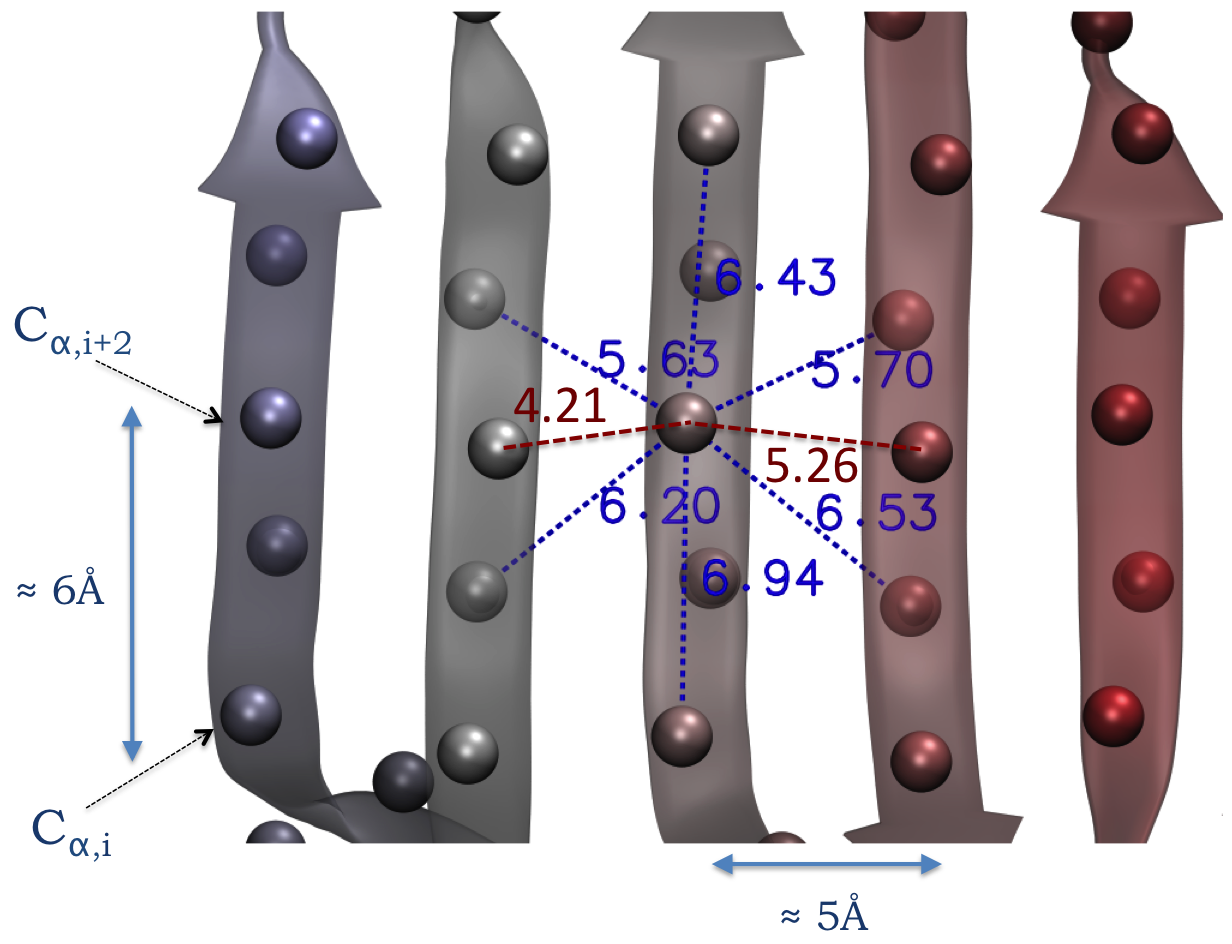}
    \caption{}\label{fig:fig13a}
  \end{subfigure}
  \begin{subfigure}{5cm}
    \includegraphics[width=\linewidth]{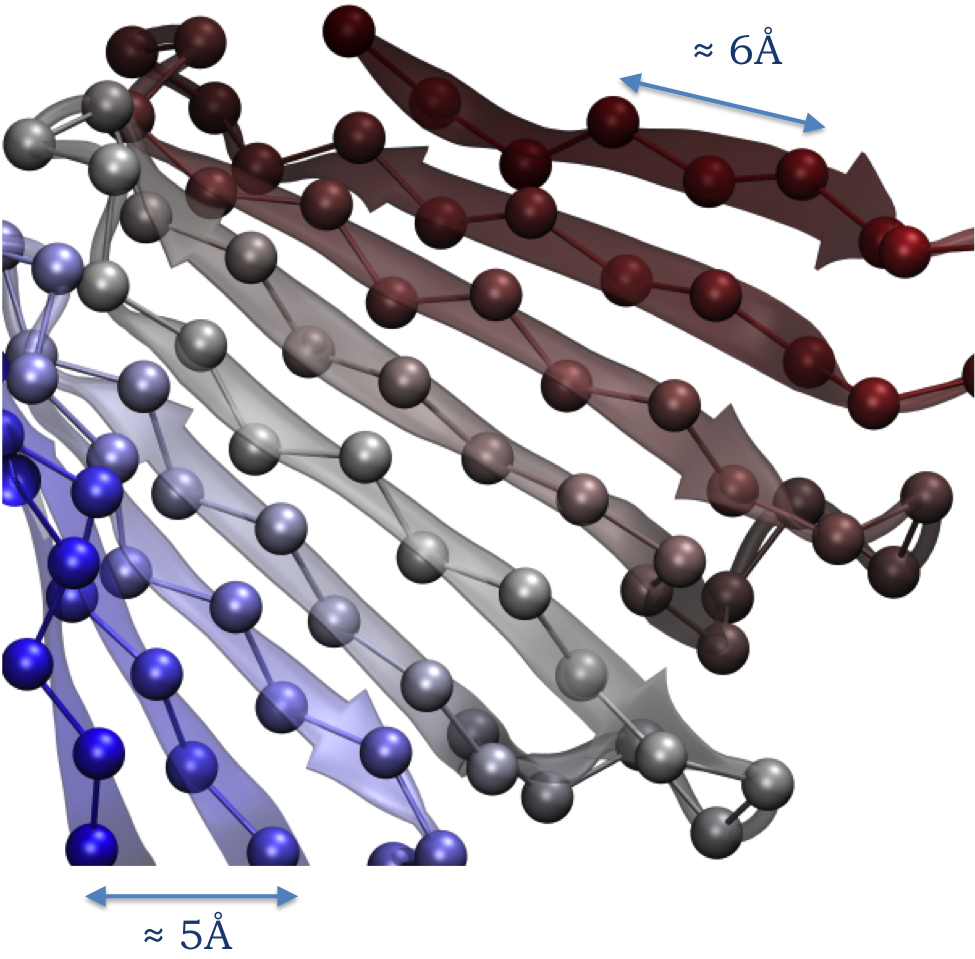}
    \caption{}\label{fig:fig13b}
  \end{subfigure}
\caption{Example of all-$\beta$ structure in proteins. (a)
A 100 residue long all-$\beta$ section of a 1OSP protein. The typical coordination number is $n_{coord}=8$ for a residue
inside a $\beta$-sheet, and neighbors contained within  $\approx 6$ {\AA}, are shown by dotted blue lines, two of which are typically hydrogen bonded (highlighted in red). Characteristic distances between adjacent strands are $\approx 5$ {\AA}, and between consecutive residues \textit{at the same height}
are $\approx 6$ {\AA}. The displayed configuration corresponds to the antiparallel case but the resulting picture is also valid for the parallel case. (b) A side view of the same configuration illustrating the characteristic in-phase arrangement of parallel strands, as well as the zig-zag conformation of consecutive residues with an angle of $\approx 120$ degrees. 
\label{fig:fig13}}
\end{figure}

\begin{figure}[htpb]
  \centering
  \captionsetup{justification=raggedright,width=\linewidth}
  \begin{subfigure}{5cm}
    \includegraphics[width=\linewidth]{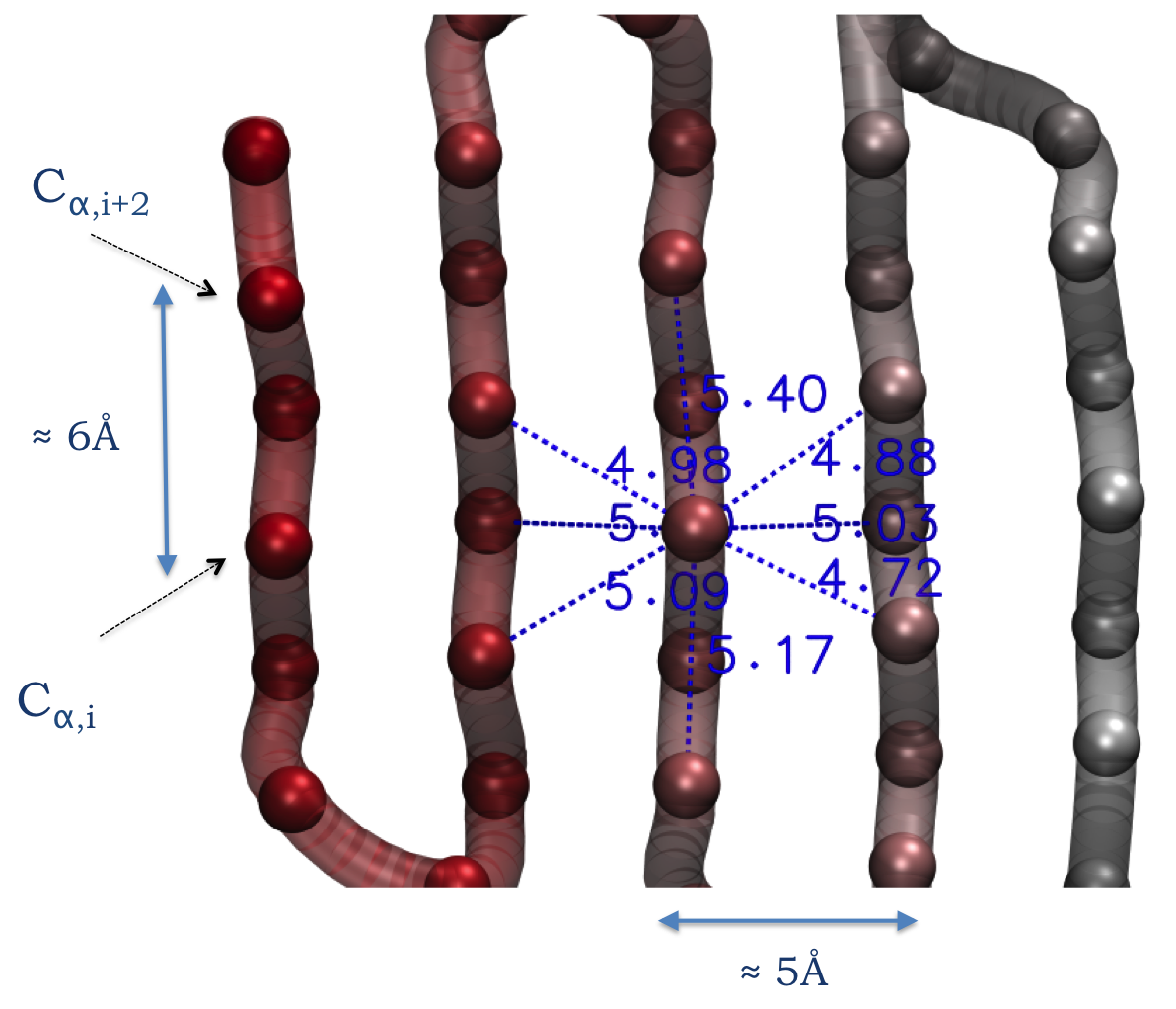}
    \caption{}\label{fig:fig14a}
  \end{subfigure}
  \begin{subfigure}{5cm}
    \includegraphics[width=\linewidth]{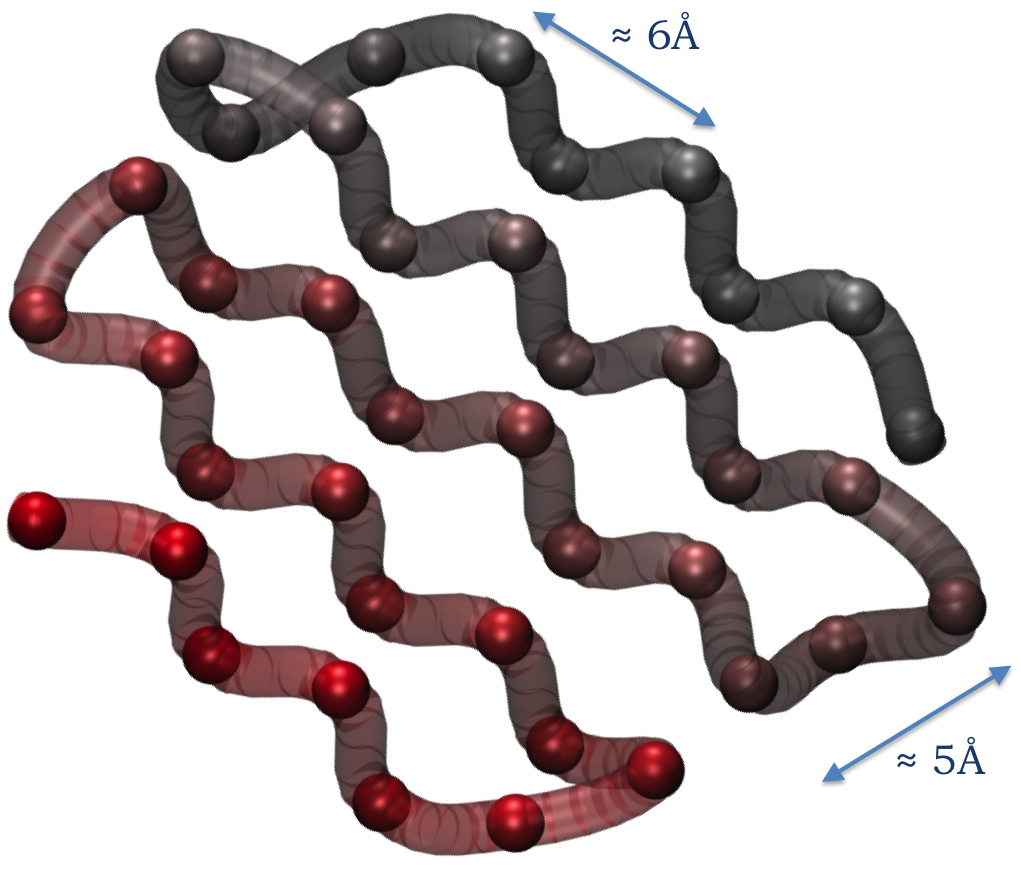}
    \caption{}\label{fig:fig14b}
  \end{subfigure}
  \caption{Example of an all-$\beta$ configuration in the elixir phase. (a) Each sphere has a 
    typical coordination number of  $n_{coord}=8$ originating from neighboring spheres within $\approx 6$ {\AA}, highlighted with blue dotted lines, in view of the accord between  the geometry of the strands in the model and in  proteins (see Figure \ref{fig:fig13}). (b) A side view of the same configuration displaying the characteristic out-of-phase arrangement of parallel strands constituting the main difference with strands in proteins. The geometrical parameters (length of the strands, intrastrand separation and zig-zag $i$-$i+2$ periodicity) are however the same in the model and for proteins. 
\label{fig:fig14}}
\end{figure}

\begin{figure}[htpb]
  \centering
  \captionsetup{justification=raggedright,width=\linewidth}
  \begin{subfigure}{5cm}
    \includegraphics[width=\linewidth]{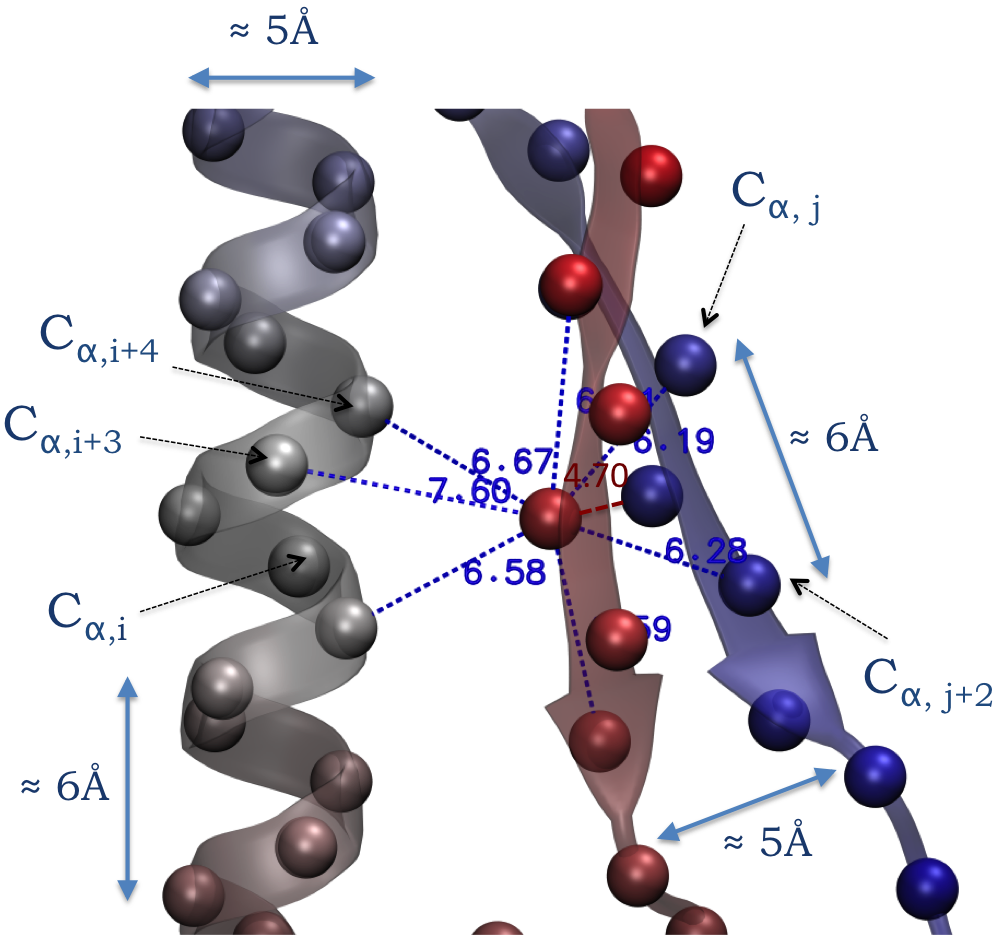}
    \caption{}\label{fig:fig15a}
  \end{subfigure}
  \begin{subfigure}{5cm}
    \includegraphics[width=\linewidth]{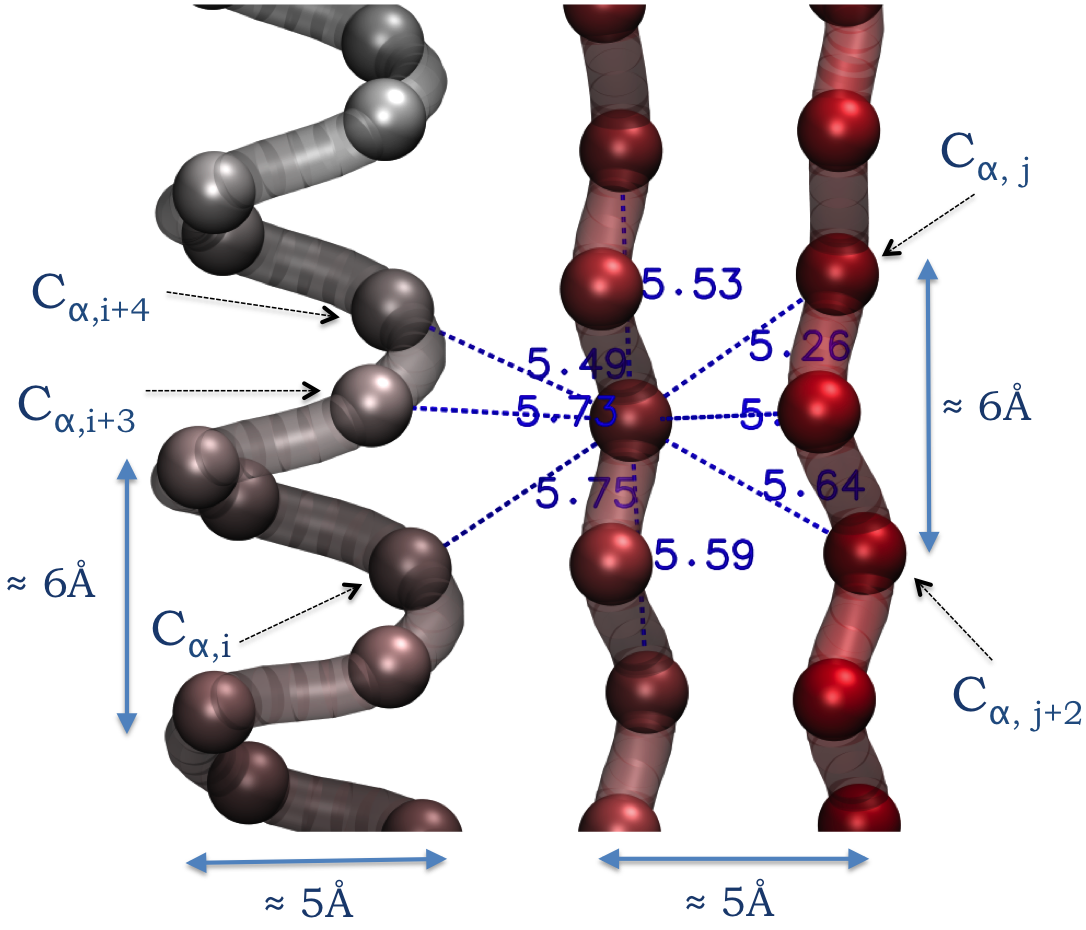}
    \caption{}\label{fig:fig15b}
  \end{subfigure}  
  \caption{Compatibility of the $\alpha$-$\beta$ configurations in a protein and in the elixir phase. (a) A protein; (b) Elixir phase.
    Typical coordination number of  $n_{coord}=8$ for a sphere embedded in the $\alpha/\beta$ environment, showing that this combined structure is comparable in energy with both all-$\alpha$ and all-$\beta$ configurations of Figures \ref{fig:fig11} -- \ref{fig:fig14}.
\label{fig:fig15}}
\end{figure}

Figure \ref{fig:fig11} shows an all-$\alpha$ conformation of a real protein (the 1ROP protein), both for each single helix (left) and the association of two helices (right).
Note that the single helix on the left is one of the two helices considered on the right. Hydrogen bonds and C$_{\alpha}^i$-C$_{\alpha}^{i+4}$ distances within $\approx 6$ {\AA} are highlighted.
Note that there are exactly 3.6 residues for each turn in a single helix of a real protein, and this results in the well-known twist in the red dotted lines (turn angle $\approx 100^{\degree}$) appearing in the left panel figure. Consider now the local environment seen by one of the residues located at the inner edge of one of the helices in the full all-$\alpha$ parallel arrangement of the 1ROP protein, as shown in the right panel of Figure \ref{fig:fig11}. As in the previous case, we have highlighted in red the two hydrogen bonds that the residue forms with the two residues one turn distant along the chain. In addition to that, however, we have also highlighted in blue all residues that are within a range of $\approx 6$ {\AA} from that residue, both within the same helix and in the parallel one. Because of the out-of-phase parallel arrangement, that residue has 3 contacts with the neighbouring parallel helix, as well as other 6 (3 above and 3 below) within the same helix, for a total of 9 possible contacts within a spherical region of diameter $\approx 6$ {\AA}. We will refer to this region as the \textit{hydrophobic box} for reasons that will become clear as we proceed. 
This bonding pattern can be compared with the counterpart in the elixir phase shown in Figure \ref{fig:fig12}. Here there are no hydrogen bonds, and hence all highlighted contacts (in blue) are those within the interaction of range $R_c \approx 6$ {\AA}, corresponding to the hydrophobic box previously alluded to.
Hence, in the elixir phase the interplay between the spherically symmetric attractive
interaction of range $R_c \approx 6$ {\AA} and the steric hindrance provided by the side spheres with diameter $\sigma_{sc} \approx 2.5$ {\AA}  (the typical value within the elixir phase) combine together to provide an effect similar to that of directional hydrogen bonds present in real proteins, for $\sigma \approx$ 5 {\AA} , the experimental value for the diameter of the van der Waals sphere of Glycine.
Outside the elixir phase, $R_c$ and $\sigma_{sc}$ do not have the correct values and the matching is no longer achieved.
Note that there are exactly 4 spheres per turn in the helical model, corresponding to a turn angle of $\approx 90^{\degree}$ resulting in the parallel pattern of the dotted lines in Figure \ref{fig:fig12} (left panel), at variance with the values of 3.6 and  $\approx 100^{\degree}$ found in real proteins (see Figure \ref{fig:fig11} left panel). Another important difference is that both right and left hand helices appear in the elixir phase, due to the achirality of the present model, at variance with real proteins. This is visible by comparing the right panels of Figures \ref{fig:fig11}, where the two helices have opposite handedness, and Figure \ref{fig:fig12} where they have the same handedness. 
A similar comparison for an all-$\beta$ conformation proves also particularly instructive.  Figure \ref{fig:fig13} displays the case of a 1OSP protein that has an all-$\beta$ native structure. Here too, hydrogen bonds with parallel strands have been highlighted in red, and neighboring residues within the hydrophobic box (size $\approx$ 6 {\AA}) have been highlighted in blue. This shows that inside a $\beta$-sheet, each residue is typically surrounded roughly by $8$ other residues within the hydrophobic box, two of which are hydrogen bonds with parallel strands. 
Again, this can be contrasted with its all-$\beta$ counterpart in the elixir phase.  Figure \ref{fig:fig14} depicts an all-$\beta$ environment within the elixir phase. Here too the combined effect of  $R_c \approx 6$ {\AA} and $\sigma_{sc} \approx 2.5$ {\AA} provides the same local arrangement for a bead in the $\beta$ conformations of real proteins of Figure \ref{fig:fig13}. Note however the different arrangement between parallel strands (in-phase in real proteins, see
Figure \ref{fig:fig13b}; out-of-phase in the elixir phase, see Figure \ref{fig:fig14b}), constituting one of the main shortcomings of the model in mimicking reality. This flaw can be corrected by incorporating a binormal-binormal interaction between the Frenet coordinate systems of main chain spheres in close spatial proximity with each other, along the lines suggested in references \cite{Banavar06,Craig06b,Bore18}. This notwithstanding, the similarity of the local environment in the elixir phase and in real proteins is striking. The slight difference in local coordination number ( $n_{coord}=8$ for an all-$\beta$ environment; $n_{coord}=9$ for an all-$\alpha$ environment) is in agreement with the statistics of the extrapolated number of contacts per beads $N_c/N$ at large $N$ discussed in Figure S4b of SI.
The above results illustrate how the geometries of the $\alpha$ and $\beta$ conformations self-tune in the elixir phase to match characteristic lengths of protein native state structures, so that
even a simple aspecific potential, combined with steric effects given by the presence of the side chain with the proper symmetry, provide an effect akin to that of directional hydrogen bonding and allows one to achieve the correct local coordination. This permits the $\alpha$ and $\beta$ motifs to coexist with each other and form combined $\alpha+\beta$ or $\alpha/\beta$ structures found in the elixir phase, as in proteins. This is displayed in Figure \ref{fig:fig15}, both for a protein and in the elixir phase.
Figure \ref{fig:fig15a} depicts the local environment seen by a residue of a real protein lying on a $\beta$ strand and having a parallel $\alpha$ helix.  As before, there are 8 neighbours within the hydrophobic cell, two of which are hydrogen bonds (highlighted in red). This can be contrasted with its counterpart within the elixir phase given in Figure \ref{fig:fig15b}, where the pattern is exactly the same, notwithstanding the absence of hydrogen bonds. 
Outside the elixir phase, this special geometrical confluence  does not occur, as shown in Figure S5 (SI), making coexistence impossible. Here the size of the side chain is too large or the range of interactions is outside the correct range, or the helix does not have the correct shape, or a combination of all these features prevent the possibility of maintaining a local environment with the correct number of contacts within the hydrophobic box. In the example given in Figure S5a (SI), the single helix has a shape different from that in the elixir phase, and when in the vicinity of a single $\beta$ strand (see Figure S5b in SI) one bond (highlighted in orange) turns out to be outside the hydrophobic box and hence matching cannot occur. The incompatibility of packing in the general case illustrates the challenge of the harmonious coexistence of the two building blocks of protein structures.
\subsection{Comparison of structural parameters with real proteins}
\label{subsec:structural}
A word of caution is in order. While all folds found in the elixir phase (and not outside it) have topologies and geometries matching those of real proteins, they are not the native states that can only be achieved incorporating the crucial information included in the sequence. Our aim here is not to try to reproduce the complexity of the protein folding mechanism, something that clearly requires more detailed approaches  \cite{Ovchinnikov17,Shaw10}. Rather, we aimed at showing how in life-as-we-know-it evolutionary biology has provided the existence of this special phase as a backdrop to build proteins that can achieve their optimal structure to perform their functions. The existence of this special phase of matter, the elixir phase, where real proteins structures may be poised to reside, allow proteins to achieve their optimal topology by using only non-specific interactions, selecting from a relatively large but limited number of possible folds. Use of the specificity of the sequence allows each protein to reach their final native state by choosing from the possible folds determined independent of sequence specificity.  As mentioned in Section \ref{subsec:future}, the existence of this phase could play an important role in the self-assembly process of several such chains.

\begin{figure}[htpb]
  \centering
  \captionsetup{justification=raggedright,width=\linewidth}
    \begin{subfigure}{5cm}
    \includegraphics[width=\linewidth]{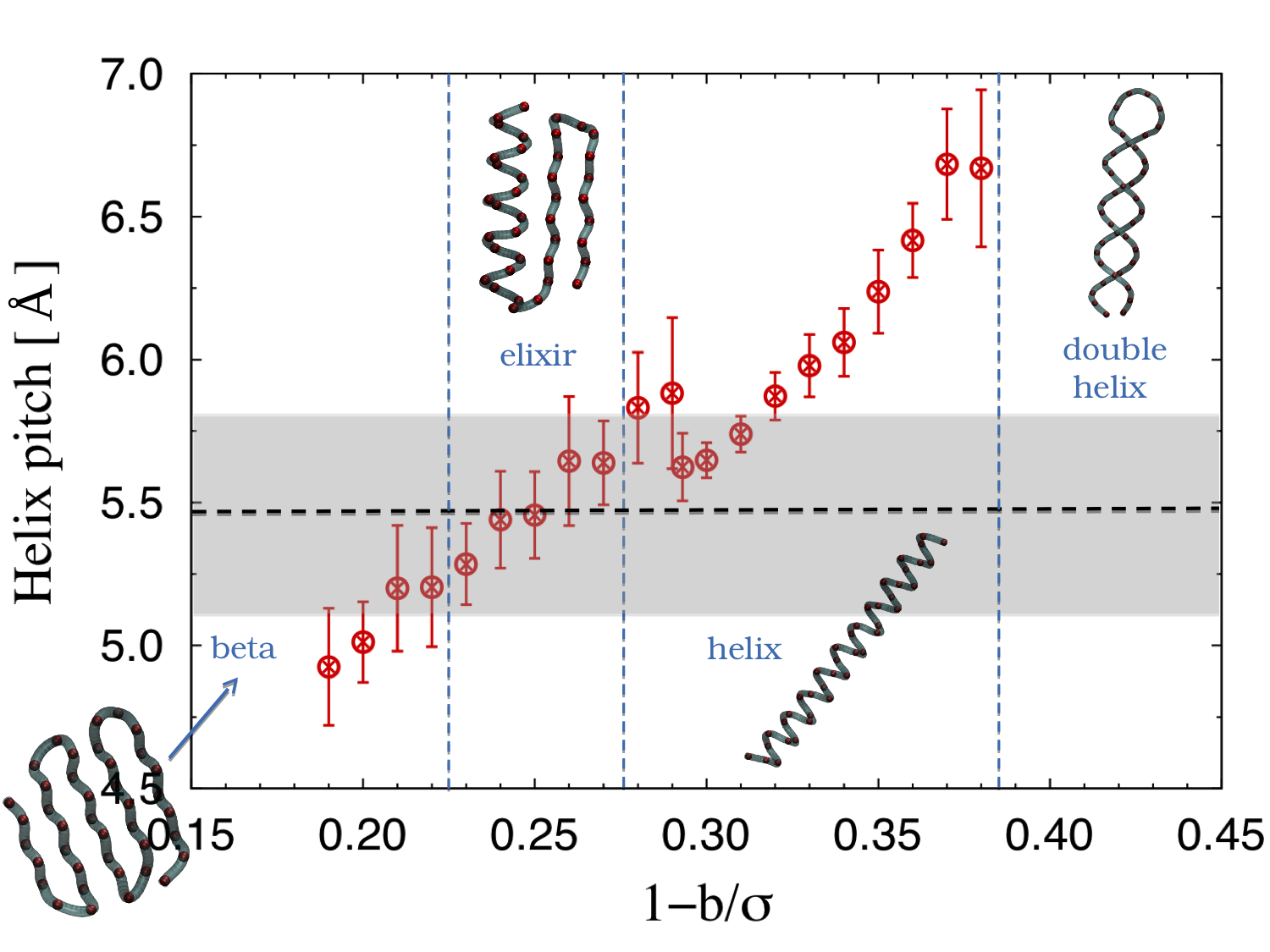}
    \caption{}\label{fig:fig16a}
  \end{subfigure}
  \begin{subfigure}{5cm}
    \includegraphics[width=\linewidth]{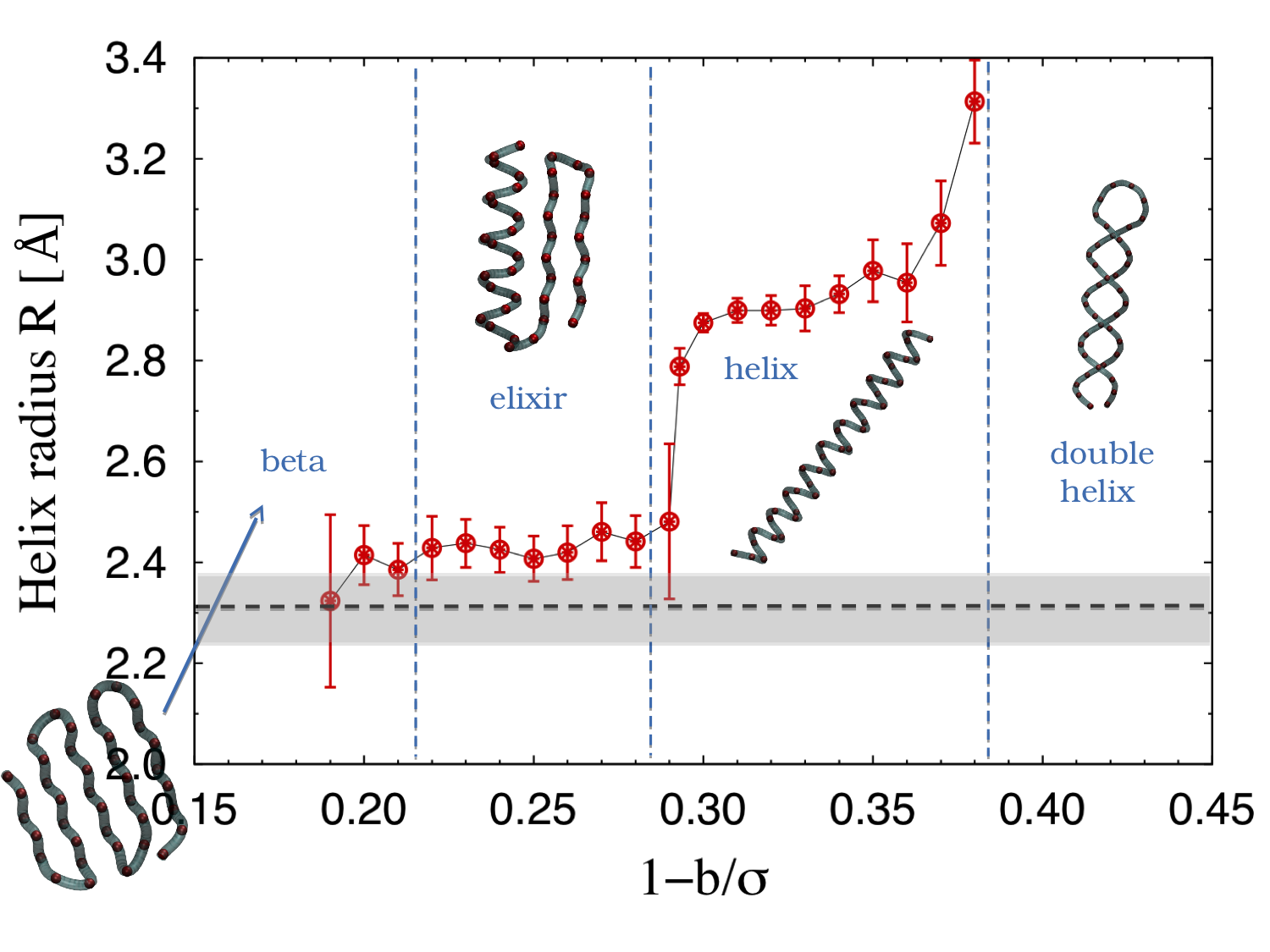}
    \caption{}\label{fig:fig16b}
  \end{subfigure}   
\begin{subfigure}{5cm}
    \includegraphics[width=\linewidth]{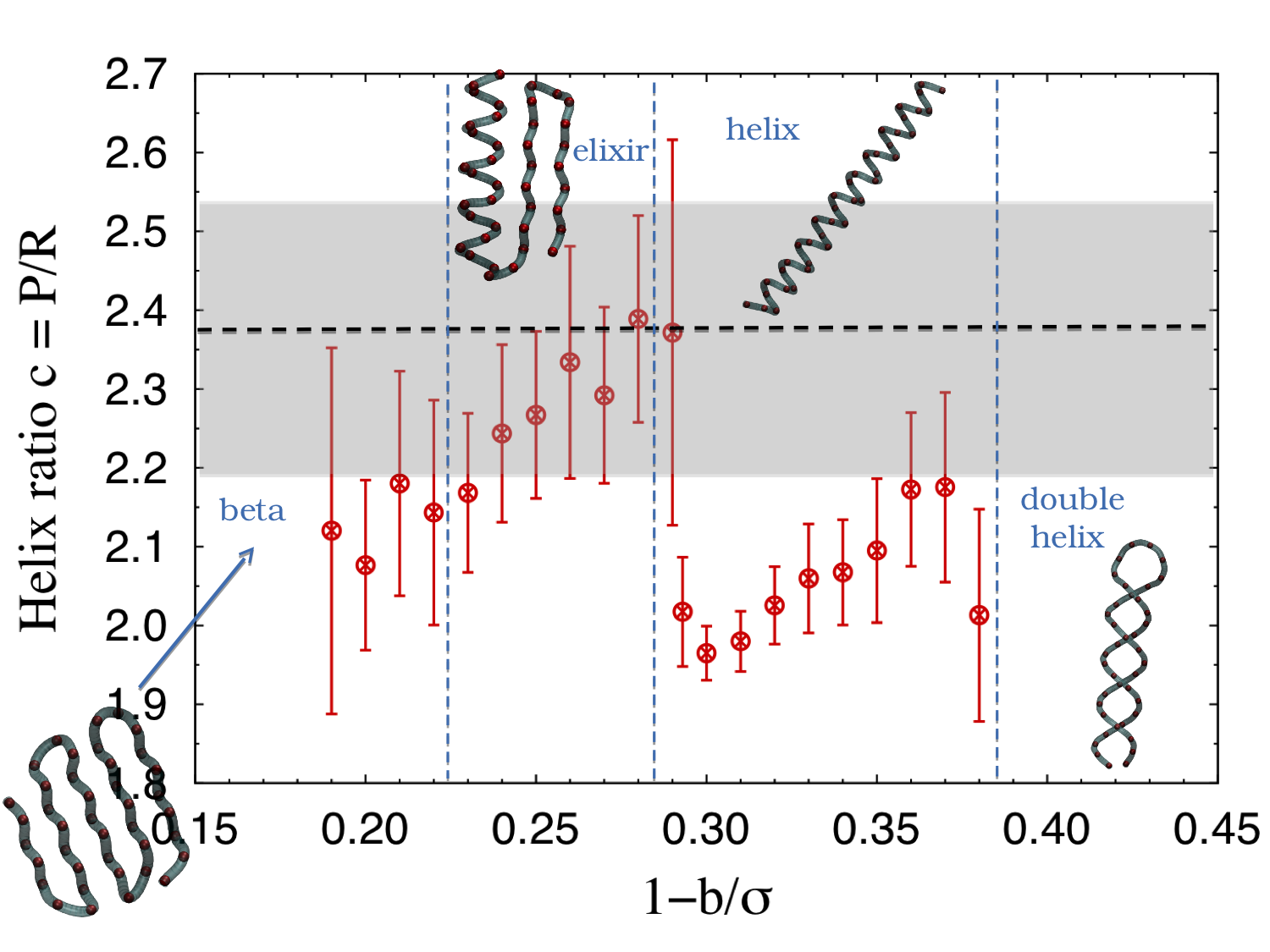}
    \caption{}\label{fig:fig16c}
\end{subfigure}
  \caption{Changes in the values of the pitch $P$ (a), radius $R$ (b), and dimensionless ratio $c=P/R$ (c) for helices crossing the elixir phase along the overlap $1-b/\sigma$ axis. Highlighted in gray are values found on the basis of a statistical analysis of proteins of all sizes \cite{Wang15}. Representative snapshots of structures in different phases are also displayed. On crossing into the elixir phase, the values found in proteins are approximately realized. We note that the increase of both the pitch and the radius in the helix (a) region arises from the need to avert steric clashes.
As $b/\sigma$ decreases (and $1-b/\sigma$ increases), the effective distances between side spheres decrease, and hence the pitch tends to increase to accommodate their packing. Note that as $R_c/\sigma$ effectively increases, the number of contacts (and hence the energy) do not increase smoothly but in a step-like fashion in view of the all-none character square-well attractive potential.
\label{fig:fig16}}
\end{figure}

How does a chain fine-tune its parameters to be poised in the elixir phase? It is interesting to compare the geometries of the structural units obtained in our model
with those found in proteins. In order to do this, we fix $b=3.81$ {\AA}, the ($C_{\alpha}^{i},C_{\alpha}^{i+1}$) distance, and rescale all other lengths in our model accordingly.
Table 1 (SI) reports the average values of the radius and pitch of a $\alpha$ helix in real proteins as obtained from a statistical analysis \cite{Cao15}, as well as the angle between $(i,i+2)$ residues in a $\beta$ strand.
As we know, the distance between the hydrogen bonded amino acids in native helices of real proteins is
$d(C_{\alpha}^i,C_{\alpha}^{i+4}) \approx 5.5$ {\AA} (see Figure \ref{fig:fig11}), corresponding to the pitch of these helices reported in Table 1 of SI.
The helix radius is $\approx 2.3 \mathrm{{\AA}}$ leading to a ratio $c=P/R \approx 2.4$. These values are compared in Figure \ref{fig:fig16} with those
obtained in our model upon increasing one of the three parameters ($1-b/\sigma$) and keeping the other two fixed to their values at the center of the elixir phase ($\sigma_{sc}/\sigma=0.5$ and $R_c/\sigma=1.167$). Each point in Figure \ref{fig:fig16} represents the result of at least 10 independent simulations, with the error bars representing the statistical fluctuations. Figure \ref{fig:fig16a} compares the helix pitch $P$ with the gray strip representing the values from real proteins as discussed in Table 1 (SI). Figures \ref{fig:fig16b} and \ref{fig:fig16c} display the analogous comparison for the helix radius $R$ and the ratio $c=P/R$. It is clear how all values within the elixir phase are compatible with those in real proteins. However, we further note that compatibility also occurs slightly outside the elixir phase, and leads to the region surrounding the elixir phase delimited by dotted lines in the phase diagrams of Figures \ref{fig:fig6} and \ref{fig:fig8}.

\begin{figure}[htpb]
  \centering
  \captionsetup{justification=raggedright,width=\linewidth}
  \includegraphics[width=0.7\linewidth]{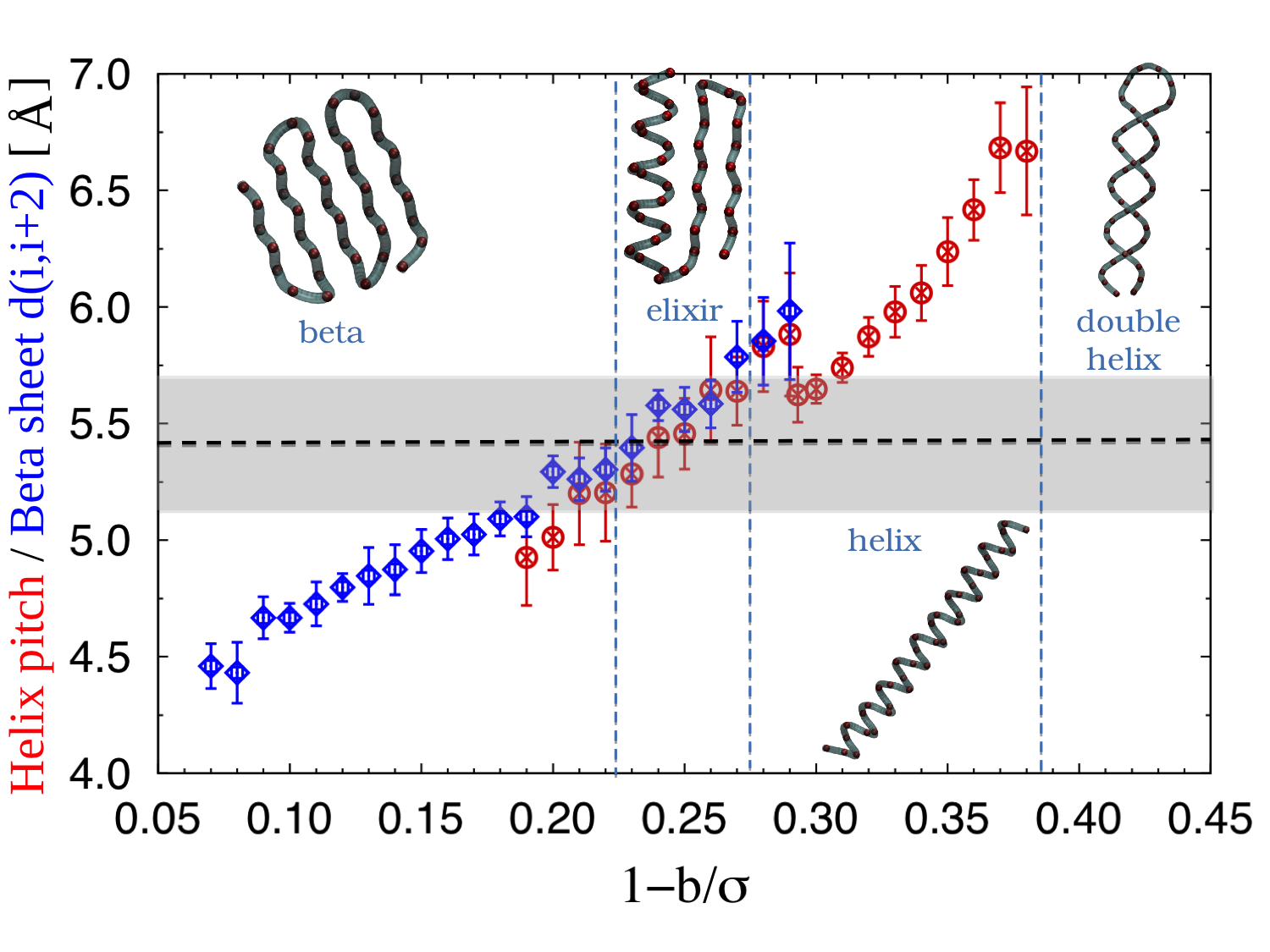} 
  \caption{Changes in the values of the distance $i$-$i+2$ of the main chain spheres in the elixir phase (and adjoining $\beta$ phase) upon increasing  the overlap $1-b/\sigma$ (Blue points). The range of values found in a statistical analysis of real proteins of all sizes is shown in gray. Representative snapshots of different ground states are also displayed. Also shown (in red) is the pitch of the $\alpha$ helix phase to highlight the coexistence of the building blocks within the elixir phase.
\label{fig:fig17}}
\end{figure}

Additional insights can be obtained by comparing the pitch of the helix in the elixir phase (i.e. the distance  $d(C_{\alpha}^{i},C_{\alpha}^{i+4})\approx 6$ {\AA} seen in Figure \ref{fig:fig12}) (right panel) and the distance $d(i,i+2) \approx 6 $ {\AA} between two next-to-consecutive beads in the strands of a $\beta$ in the elixir phase (see Figure \ref{fig:fig13}). This is
reported in Figure \ref{fig:fig17} as a function of $1-b/\sigma$, showing once more that only in the elixir phase (and in fact also slightly outside it) compatibility occurs and
$\alpha/\beta$ and $\alpha+\beta$ structures can be found.

Interestingly, this result is in perfect accord with recent suggestions given in Ref. \cite{Cardelli17}, that a peak in the radial distribution function at $\approx 6 $ {\AA} is indeed a distinct feature of protein-like folds.
Two futher points are worth noting. First, in the entire $b/\sigma$ range of the elixir phase ($\approx 0.7-0.8$ or equivalently the overlap $(1-b/\sigma) \approx 0.2-0.3)$ and using $b=3.81$ {\AA},  the values of $R_c$  characteristic of the elixir phase 
are found to be in the range 5.6-6.3 {\AA}, that coincide with the hydrogen bond length derived from quantum chemistry. Second,
there is a significant difference between the ground states in the elixir phase and native states of protein structures in the way contacts occur. In our model structures, the most frequent contacts occur between $i$ and $i+2$ beads and those play a significant role in driving the folding of the chain. This is at odds with actual protein structures where favorable interactions between $i$ and $i+2$ residues are infrequent. 
\subsection{Discussion and Future perspectives}
\label{subsec:future}
The elixir phase emerges as a new phase due to the elimination of the cylindrical symmetry that resulted in the marginally compact phase. By a harmonious combination of non-specific attraction between non-consecutive beads in the chain, and of the excluded volume effect provided by the presence of the side chain, the system is able to find a ground state where the ground state conformations have remarkable similarity with those obtained via a very different path in real proteins. In essence, the elixir phase is able to reproduce the same delicate balance found in real proteins as a result of many different non-covalent interactions, using a much reduced set of parameters and ingredients. 

\begin{figure}[htbp]
  \centering
  \captionsetup{justification=raggedright,width=\linewidth}
\includegraphics[width=0.7\linewidth]{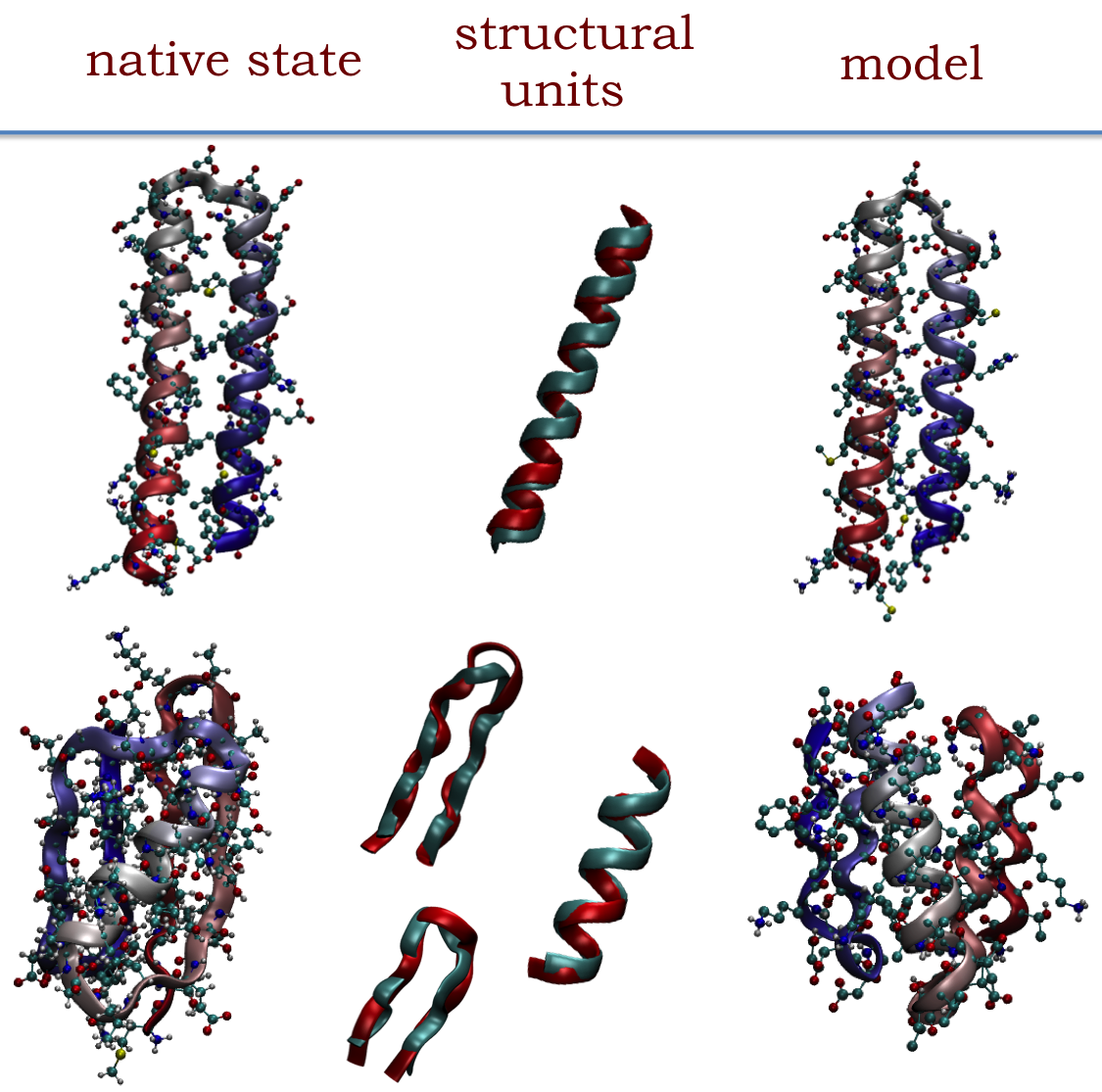}
 \caption{Examples of comparison between ground state structures in the elixir phase and the native folds of proteins comprised of $N=56$ amino acids. For the model, the sequence information was incorporated after the ground state was determined. (Top) ColE1 rop protein (PDBid 1ROP), whose native state is formed by two $\alpha$ helices connected by an unstructured strand (left), and the corresponding ground state in the model with parameters: $\sigma = 5$ {\AA}, $1-b/\sigma=0.28$, $\sigma_{sc}/\sigma=0.533$ and $R_c/\sigma = 1.167$. The root-mean-square deviation (RMSD) between the native state of the protein and the ground state of the model is $\approx 4.1$ {\AA}. (Bottom) Protein G (PDBid 3GB1), whose native state is formed by two $\beta$ antiparallel hairpins connected by a single $\alpha$ helix (left) and the topologically similar ground state in the model with parameters: $\sigma = 5 $ {\AA}, $1-b/\sigma=0.25$, $\sigma_{sc}/\sigma=0.416$ and $R_c/\sigma = 1.167$. The root-mean-square deviation (RMSD) between the native state of the protein and the ground state of the model is $\approx 7$ {\AA}. The central panel shows the fidelity of the overlap of the building blocks of the protein structures ($\alpha$ helices and $\beta$ hairpins) to those in the elixir phase (RMSD) $\le 2.0 $ {\AA}. Structural units of real proteins are shown in red, those from the model are in cyan. 
\label{fig:fig18}}
\end{figure}
Let us consider a specific example. Consider protein ColE1 rop protein (PDBid 1ROP), whose native state has the all-$\alpha$ conformation shown in Figure \ref{fig:fig18} (top left panel). We can now scan the elixir phase ground states (see Figure S6 in SI for a representative gallery of the structures found in the elixir phase and in the immediate region surrounding it where the single units ($\alpha$ helix and $\beta$ sheet) have structural parameters matching those found in real proteins. This is the region delimited by dotted lines in the phase diagrams of Figures \ref{fig:fig6} and \ref{fig:fig8} that include the elixir phase. An all-$\alpha$ helix is clearly visible in the bottom right part of the gallery of Figure S6 (SI). Clearly, this includes only the $C_{\alpha}-C_{\beta}$ beads along the chain. However it is possible to reconstruct the same sequence of protein 1ROP using the PULCHRA tool \cite{Rotkiewicz08} that reconstructs the full-atom protein model starting from its coarse-grained representation. The result is shown in the top right panel of Figure \ref{fig:fig18}. As expected, the similarity is striking. One important difference can be seen when comparing the details of the two parallel helices, that have opposite handedness in protein 1ROP (top left) and same handedness in the elixir ground state (top right). Yet, each single unit is essentially identical, as shown again in the top central panel of Figure \ref{fig:fig18} where the real and the model helices are superimposed to one another. Many other examples with different topologies and number of residues can be found. The bottom left panel of Figure \ref{fig:fig18} shows Protein G (PDBid 3GB1) as an example of an $\alpha/\beta$ conformation. Again, it is possible to find an elixir ground state having a similar topology (see second right in the central row of Figure S6 in SI), leading to the reconstructed artificial protein depicted in Figure \ref{fig:fig18} bottom right panel. Even in this case, the central panel at the bottom shows the superimposition of the real and the model units that in this case include also $\beta$ strands.

\begin{figure}[htpb]
  \centering
  \captionsetup{justification=raggedright,width=\linewidth}
  \begin{subfigure}{7cm}
    \includegraphics[width=\linewidth]{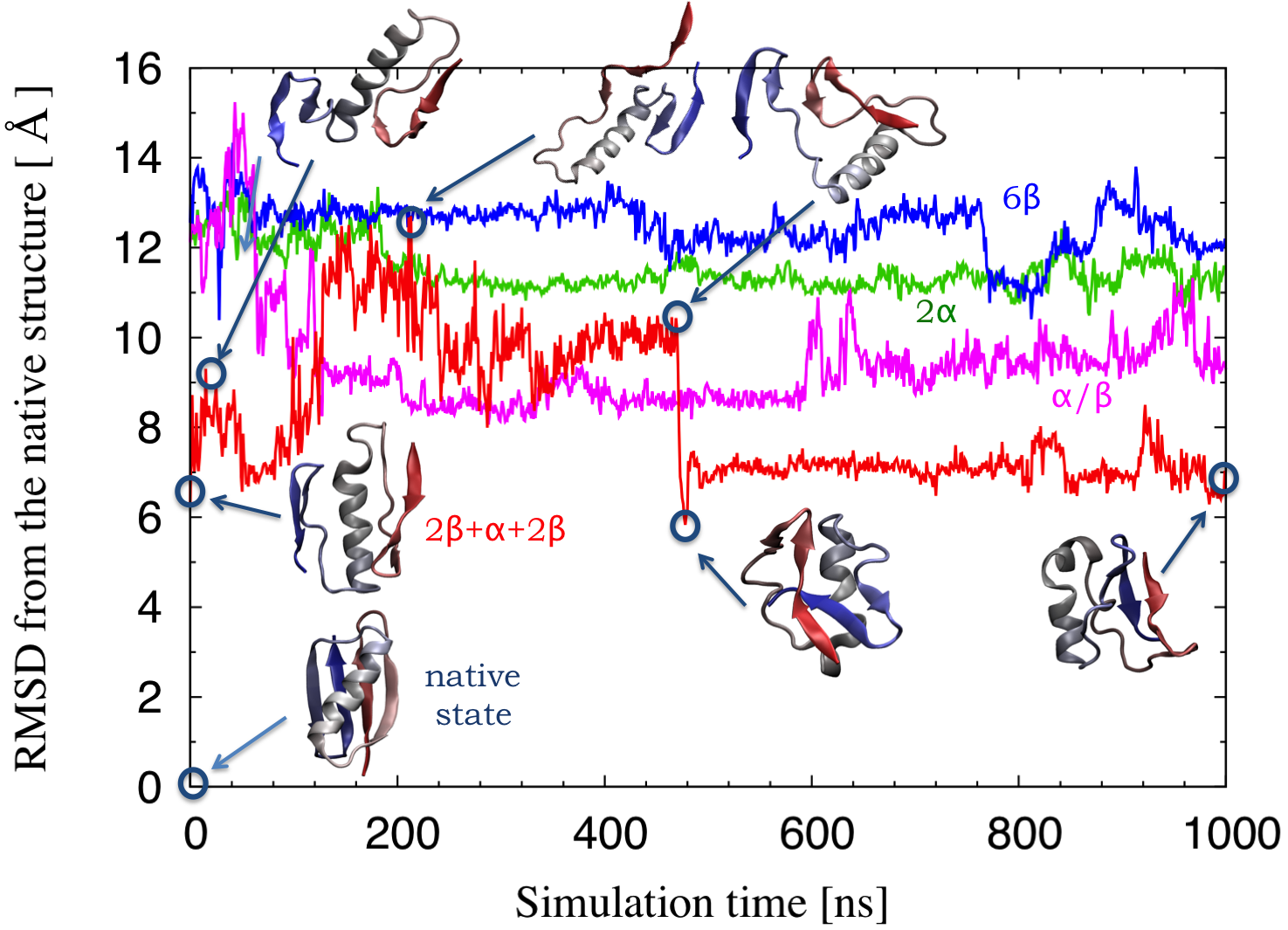}
    \caption{}\label{fig:fig19a}
  \end{subfigure}
  \begin{subfigure}{7cm}
    \includegraphics[width=\linewidth]{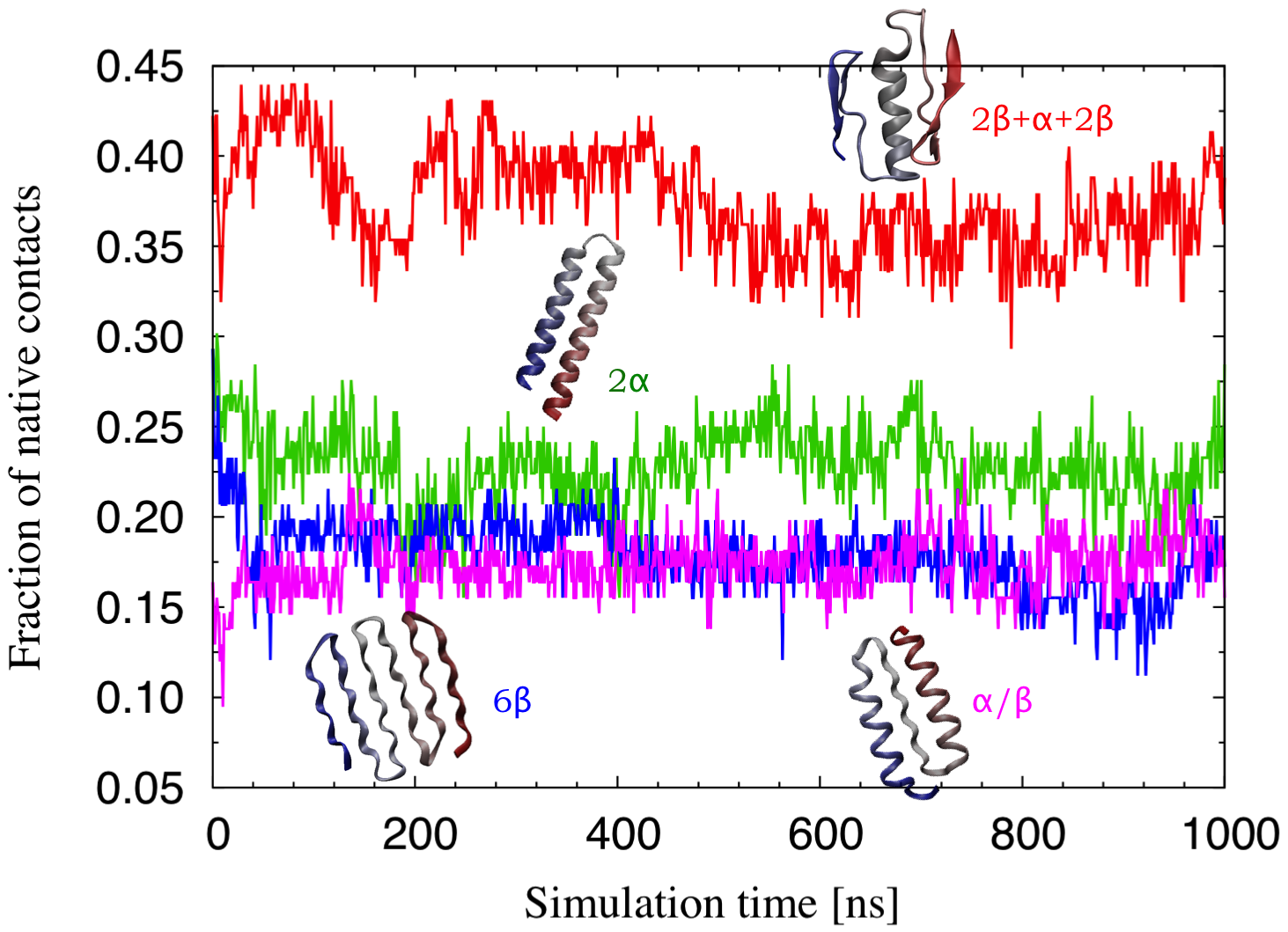}
    \caption{}\label{fig:fig19b}
  \end{subfigure}   
  \caption{(a) Comparison of the Root-mean-square-deviation (RMSD) from the native conformation of Protein G(PDBid 3GB1) starting from the $2\beta+\alpha+2\beta$ ground state obtained in Figure \ref{fig:fig18}, as well as from three other alternative decoys: the 6$\beta$, the 2$\alpha$, and the $\alpha/\beta$. In all cases, the ground state was  dressed with full atomistic details from the specific sequence of Protein G(PDBid 3GB1), and a isobaric-isothermal molecular dynamics simulations was performed in explicit water using the GROMACS package up to $2 \mu$ sec. Snapshots corresponding to relevant times of the trajectory of the $2\beta+\alpha+2\beta$ evolution are also displayed. (b) Same calculation comparing the fraction of native contacts. Here, snapshots of all original four ground states structures are also depicted as insets.   
\label{fig:fig19}}
\end{figure}

The two examples given in Figure \ref{fig:fig18} compare the \textit{exact} ground state of the elixir model with the native folds of two real proteins, and show that the elixir phase includes conformations with topologies similar to the native ones, albeit with some differences. This similarity suggests that these elixir ground states could also be regarded as approximate molten globule phase \cite{Baldwin13} of the corresponding real proteins, where the general topologies has already self-assembled but the native state is still to be reached. Unlike the coil state, the ground state obtained here is at the bottom of the funnel landscape \cite{Onuchic04}, and constitutes then a much more convenient starting point toward a more detailed calculation. This idea is illustrated in     
Figure \ref{fig:fig19}, where we consider the $2\beta+\alpha+2\beta$ ground state reported in the Figure \ref{fig:fig18}
as having the overall topology very close to the actual native state of Protein G (PDBid 3GB1), as well as three other different ground states (decoys): the 6$\beta$, the 2$\alpha$, and the $\alpha/\beta$ having the same sequence pertaining to Protein G (PDBid 3GB1). We then performed unconstrained molecular dynamics simulations in explicit solvent up to
$2 \mu$ sec using the GROMACS package \cite{Berensen95,Hess08} and compared the Root-mean-square-deviation (RMSD) (a), as well as the fraction of native states (b), from Protein G(PDBid 3GB1) native state. In both cases, the optimal value is achieved for the $2\beta+\alpha+2\beta$ decoy, as expected. Note that the RMSD does not eventually go to a lower value, due to the frustrating effect played by the shortcomings of the elixir ground states discussed in Figures \ref{fig:fig11}--\ref{fig:fig14}. A possible remedy to this drawback might stem from the addition of binormal-binormal adjustments between the Frenet reference frames of main chain spheres in proximity to each other, as recently discussed in Refs. \cite{Banavar06,Craig06b,Bore18}.

The ordering trend of the other transitions agrees with those found in real proteins. This can be best represented in terms of the contact order parameter \cite{Plaxco98,Bonneau02}
\begin{eqnarray}
  \label{sec3:eq1}
  CO &=& \frac{1}{N N_c} \sum_{j>i+1}^{N_c} \Delta S_{ij}
\end{eqnarray}
where $N_c$ is the total number of contacts and $N$ is the number of residues (beads) in the chain and the sum is over pairs of amino acids in contact. Here $\Delta S_{ij}$ is the sequence separation between $i$ and $j$ residues.

The contact order $CO$ has been shown \cite{Plaxco98} to be $\approx 10 \%$ in helical proteins, and $\approx 17 \%$ in $\beta$-sheet proteins.
The corresponding folding temperatures were found to be $\approx 30^{\circ}$C and $\approx 25^{\circ}$C in the two cases, respectively. Both trends can be rationalized in terms of the entropic advantage of breaking non-local contacts (occurring in $\beta$ sheets) as compared to local contacts (occurring in $\alpha$ helices). Hence a lower temperature is sufficient to have the same contribution to the free energy difference for a larger entropy change.

Notwithstanding the aforementioned differences with real proteins, the same trend is observed in our model, with the transition temperature for the single $\beta$ conformation ($k_BT/\epsilon \approx 0.4$, Figure \ref{fig:fig3b}) being lower than that of the single helix conformation ($k_BT/\epsilon \approx 0.6$ Figure \ref{fig:fig3a}). A measure of the contact order parameter (\ref{sec3:eq1}) gives $CO \approx 16 \%$ in the single $\beta$ and $CO \approx 9 \%$ in the single helix cases. Interestingly, the average contact order found in the elixir phase is found to be $CO \approx 16 \%$ closer to the single $\beta$ than the single helix phases, similar to that in real proteins (see Table 1 in \cite{Plaxco98}).

While interesting on its own right, the  existence of the elixir phase is relevant to the self-assembly of many such chains. The present model is greatly simplified and more ingredients will need to be added to understand the detailed behaviour of proteins \cite{Shaw10,Ovchinnikov17}. However, this is not the case when it comes to \textit{de novo} protein design through a self-assembly process of short peptides \cite{Huang16,Ljubetic17,Shen18,Li19,Bera19}. Indeed, such self-assembly can be facilitated by exploiting general physical and symmetry principles such as those presented here.

A simple example in the framework of amyloid formation \cite{Auer07a} might be useful to highlight this point. Consider a short chain (say comprised of 20 monomers) with parameters such that the ground (folded) state of the chain is a single helix. From Figure \ref{fig:fig8a}, we know that it is possible to control the parameters to guarantee that this is the case. When a sufficiently large number of such chains are assembled together in a given volume, there is a known strong tendency to form $\beta$-sheet assemblies via a nucleation-and-growth mechanism \cite{Bera19} that can be captured by a model like ours that does not require a detailed knowledge of the specific chemical interactions \cite{Auer07a}, Work along these lines is underway and will be presented elsewhere.

\section{Conclusions}
\label{sec:conclusions}
In this paper, we discussed how a simple modification of a conventional homopolymer model provides a very rich and informative protein-like phase diagram. The model relies on a two-beads representation of each amino-acid, one for the backbone and one of the side chain. In addition to excluded volume interaction, a short range attraction between non-consecutive backbone beads is enforced to mimic the hydrophobic interactions as in conventional polymers. Unlike conventional polymers, consecutive backbone beads are allowed to partially interpenetrate thus resulting in the removal of the original spurious spherical symmetry in favour of a uniaxial, cylindrical, symmetry. As in liquid crystals, this symmetry breaking opens an intermediate phase, the marginally compact phase, where conformations formed by single helices and single $\beta$-strands are found in distinct regions of parameter space.
The addition of a single side chain bead located in a specific direction (the negative normal in the Frenet frame) in a plane perpendicular to the chain axis further reduces the symmetry from uniaxial to biaxial and results in the onset of the elixir phase, where multiple folds have essentially the same energy and one can switch from one to another due to thermal fluctuations.

In term of comparison with real proteins, the elixir phase is centred at values (3.81 {\AA}   for the C$_{\alpha}$-C$_{\alpha}$ distance, 2.5 {\AA}  for the diameter of the average van der Waals sphere associated with side chains, and 6 {\AA}  for the range of the hydrophobic box) that are characteristic of real proteins and emerge as an output of the calculation, rather than as an input as in many coarse-grained models similar in spirit to the present one. Recall that our calculations explored the phase diagram for the model for all sets of parameters in an unprejudiced manner. The origin of this can be traced back to the fact that, only within this elixir phase, the helices and the $\beta$-strands acquire geometries matching those occurring in real proteins and are able to compete in energy.
In addition to the important applications that this can have in terms of conformationally based nanomachines relying on switching from one conformation to another, the existence of the elixir phase unambiguously shows that a protein can fold to a limited number of possible conformations belonging to one of the four paradigmatic classes (all-$\alpha$, all-$\beta$, $\alpha/\beta$, $\alpha+\beta$), driven by general considerations of geometry and the absence of spurious symmetries that reduce the conformational entropy in a way akin to that given by directional interactions, hence mimicking the presence of hydrogen bonds.

In the elixir phase, all four characteristic topologies (all-$\alpha$, all-$\beta$, $\alpha/\beta$, and $\alpha+\beta$) found in real proteins are present. The elixir phase is surrounded by other phases, the single helix, the single $\beta$, the coil, and the globule, where a single unique ground state is found for each specific state point (i.e. for specific values of the parameters). In the single helix phase, the radius and the pitch of the helix smoothly changes on changing the parameters, still preserving the single helix character, and the helix geometry is quite distinct from that of a standard protein helix. Likewise, the single $\beta$ phase has the same topology as the all-$\beta$ phase found in the elixir phase but with incorrect geometrical parameters (the $i$-$i+2$ angle, and the length of each strand). Both units self-tune themselves to the correct values only when the parameters have values such that they are located within the elixir phase. By doing this, they are able to compete in energy and to form combined structures.

There are three lessons that we can learn from these findings. The first lesson is related to the minimal representation of a protein within a homopolymer coarse-grained model. Our results indicate that a C$_{\alpha}$-C$_{\beta}$ two beads representation of each residue is required, a single  C$_{\alpha}$ bead representation, frequently encountered in the literature, being not sufficient. The introduction of the C$_{\beta}$ is necessary to provide the model with the broken symmetry, and this drastically reduces the conformational entropy of the chain from an astronomically large number characteristic of the glassy ground state of globule, to a finite (albeit large) number of local minima out of which the sequence can select the final native structure. This is not the only possible way of achieving this task. Other possibilities are the introduction of directional interactions (e.g. hydrogen bonds), as well as the introduction of sequence heterogeneity (in the form, for instance, of hydrophobic-hydrophilic character). They are not mutually exclusive  --indeed natural proteins exploit a harmonious combinations of all of them, and this explain why models with different ingredients are able to display similar folds. This is lesson two from our results. The final lesson that can be inferred from our results is related to the conventional view of the funnel hypothesis in the energy landscape theory that ascribes the capability of avoiding kinetic trapping occurring in the pathway toward the native state uniquely to the sequence. We showed that it is possible to progress considerably down in the funnel without any information regarding the sequence. The correct symmetry as well as directional interactions (i.e. hydrogen bonds) are two other ways to achieve the same goal. This view is also shared in recent studies \cite{Cardelli17} that explore the designability of protein structures. It would be very interesting to perform a kinetic study of our model to see whether the elixir phase is reached through the assembly of small cooperative units, as the present study seems to suggest, and as expected in the so-called foldon representation \cite{Englander17,Baldwin17}.
However, the greatest interest in a coarse-grained model as that presented here stems from applications. The remarkable simplicity of the model, combined with its capability of predicting protein-like folds, may pave the way for controlled colloidal experiments \cite{Zeravcic17,Zerrouki08} and artificial self-assembly processes \cite{Dodd18}. In particular, the multichain version of the present model could be used in designing and guiding self-assembling peptide-based processes that are currently of great interest for its biomedical applications \cite{Huang16,Ljubetic17,Shen18,Li19,Bera19}.

\acknowledgments{
  We are indebted to Michele Cascella, Ivan Coluzza, Brian Matthew, Flavio Romano, George Rose, Francesco Sciortino, Luca Tubiana, and Pete von Hippel for useful discussions. The use of the SCSCF multiprocessor cluster at  the Universit\`{a} Ca' Foscari Venezia and of the high performance computer Talapas  at the University of Oregon is gratefully acknowledged. The work was supported by MIUR PRIN-COFIN2017 \textit{Soft Adaptive Networks} grant 2017Z55KCW (A.G), a G-1-00005 Fulbright and University of Oregon Research Scholarship (T.S), the Vietnam National Foundation for Science and Technology Development (NAFOSTED) under Grant No. 103.01-2016.61} (T.X.H.), an excellence project 2017 of the Cariparo Foundation (A.M.).





\end{document}